\documentclass[traditabstract]{aa}
\usepackage[utf8]{inputenc}
\usepackage{setspace}
\usepackage{amsmath}
\usepackage{amssymb}
\usepackage{microtype}
\usepackage{natbib}
\usepackage{graphicx}
\usepackage{txfonts}
\usepackage{color}
\usepackage{multirow}
\usepackage{xspace}
\bibpunct{(}{)}{;}{a}{}{,} 

\newcommand{\src}{HESS~J1731$-$347\xspace}

\newcommand{\xmm}{\emph{XMM-Newton}\xspace}

\newcommand{\escm}{\,erg\,s$^{-1}$\,cm$^{-2}$\xspace}
\begin{document}

\title{\xmm observations of the non-thermal supernova remnant \src (G353.6-0.7)}
\author{V.\,Doroshenko\inst{1} \and G.\,Pühlhofer\inst{1} \and A.\,Bamba\inst{2,3} \and F.\,Acero\inst{4} \and W.\,W.\,Tian\inst{5,6,7} \and D.~\,Klochkov\inst{1}\and A.\,Santangelo\inst{1}}
\institute{Institut für Astronomie und Astrophysik, Sand 1, 72076 Tübingen, Germany, \email{doroshv@astro.uni-tuebingen.de} \and
Department of Physics, The University of Tokyo, 7-3-1 Hongo, Bunkyo-ku, Tokyo 113-0033, Japan \and
Research Center for the Early Universe, School of Science, The University of Tokyo, 7-3-1 Hongo, Bunkyo-ku, Tokyo 113-0033, Japan \and
Laboratoire AIM, IRFU/SAp – CEA/DRF – CNRS – Universit\'e Paris Diderot, B\^at. 709, CEA-Saclay, Gif-sur-Yvette Cedex, France \and
Key Laboratory of Optical Astronomy, National Astronomical Observatories, Chinese Academy of Sciences, Beijing 100012, China \and
Institute of Astronomy and Department of Physics, National Tsing Hua University, Hsinchu, Taiwan \and 
Department of Physics and Astronomy, University of Calgary, Calgary, Alberta T2N 1N4, Canada.}

\bibliographystyle{aa}

\abstract{We report on the analysis of \xmm observations of the  non-thermal shell-type supernova remnant \src (G353.6-0.7). For the first time the complete
remnant shell has been covered in X-rays, which allowed direct comparison with
radio and TeV observations. We carried out a spatially resolved spectral
analysis of \xmm data and confirmed the previously reported non-thermal power-law X-ray spectrum of the source with negligible variations of spectral index
across the shell. On the other hand, the X-ray absorption column is strongly
variable and correlates with the CO emission thus confirming that the
absorbing material must be in the foreground and reinforcing the previously
suggested lower limit on distance. Finally, we find that the X-ray emission of
the remnant is suppressed towards the Galactic plane, which points to lower shock
velocities in this region, likely due to the interaction of the shock with the
nearby molecular cloud. 
}

 \authorrunning{V. Doroshenko et al.}
\maketitle
\section{Introduction} \src (G353.6-0.7) belongs to a small group of supernova remnants
(SNR) with associated TeV and non-thermal X-ray synchrotron emission from the
expanding supernova shock fronts. The observed X-ray emission is evidence of
relativistic TeV electrons that are accelerated in the SNR shocks. The TeV
emission from \src could stem from relativistic electrons (through
inverse Compton emission) or, alternatively, from protons (through
$\pi^0$-decay). In order to disentangle the two processes and derive
corresponding energy budgets, the parameters of the environment into which the
shocks are propagating need to be known. The progenitor of HESS~J1731-347 must have been a massive star
as its central compact object (CCO), i.e. a neutron star is detected in its centre.
Therefore, the remnant is likely surrounded by and is possibly expanding into an area of high molecular
density. However, in order to achieve its current size ($\sim30$\,pc diameter
assuming a distance of $\sim3.2$\,kpc to the source) while still maintaining a
high shock speed, the forward shock must have propagated for a large fraction
of the remnant's lifetime in a tenuous medium. A likely explanation is that the
remnant is expanding inside the stellar wind bubble blown by the progenitor
star \citep{Cui16}. The same scenario has also been  invoked to explain the size
of RX J1713.7-3946 \citep{Berezhko08}.

The distance to \src is not firmly known up to date. It is likely
that the remnant is located in one of the three Galactic spiral arms crossing
the line of sight at the near side of the Galactic centre:  in the
Scutum-Crux arm (at a distance of $\sim3$\,kpc), in  the Norma arm ($\sim4.5$\,kpc), or in the 3~kpc
arm at 5-6\,kpc distance. A location beyond the Galactic centre distance would imply
a very high TeV luminosity and seems unlikely. A reasonable lower limit of 3.2~kpc
comes from a gradient of the X-ray absorption column across the source matching
the integrated gas density up to that distance, as derived from $^{12}CO$ and HI
data \citep{discovery_paper}. The remnant could be located on the far side of
the Scutum-Crux arm, at a similar distance  (in projection) of a nearby HII
region at $\sim$3.2\,kpc distance \citep{Tian08}; however, no morphological or
other obvious link exists between the two sources. If \src were
located at this distance, it could be co-located with the bulk of absorbing
material (i.e. inside a molecular cloud region) that is located in the
Scutum-Crux arm, and possibly interact with the material. However, from the
existing data, no evidence for such an interaction has been found so far.

A near distance (3.2\,kpc to 4.5\,kpc) is also favoured  by the analysis of the
X-ray spectrum of the central neutron star \citep{Klochkov13,Klochkov15}.
Alternatively, it has been argued that \src could be located in the
3\,kpc arm \citep{Fukuda14}. While the TeV profile on its own is statistically
compatible with a flat profile \citep{discovery_paper}, \cite{Fukuda14} have
argued for a partial correlation of the TeV profile with an azimuthal gas
profile in the projected direction towards the SNR in the 3\,kpc arm distance
range. The TeV emission has been interpreted in this scenario by a blend of
hadronic emission (correlated with the gas density) and leptonic emission.
\cite{Fukuda14} also favoured this scenario based on the spatial
distribution of the surrounding molecular clouds.
Finally, using the roughly flat TeV azimuthal profile, \cite{Nayana17} argued
that the remnant is likely not expanding in an anisotropic medium as would be
expected from molecular clouds. They then used indications for an
anti-correlation with 325~MHz radio synchrotron emission to argue that the TeV
emission is due to electrons, explaining the differences as being due to
different magnetic field strengths and correspondingly different synchrotron
burnoff energies of the radiating electrons.

\cite{Cui16} have shown that within a framework to model the SNR evolution and
associated particle acceleration, it is difficult to obtain a hadronic TeV
emission scenario for \src independent of the actual distance to the
SNR. From the upper limit in the GeV band derived with Fermi-LAT
\citep{Acero15} it was argued as well that the TeV emission from \src
should be leptonically dominated. However, the same arguments that have been
used by \cite{Gabici14} to model the GeV-TeV emission of
RX~J1713.7-3946 as hadronic emission could also be applicable to HESS
J1731-347. The steep GeV spectrum on its own therefore does not provide a
definite discrimination between the emission models.

X-ray data of \src can be used to probe the surrounding geometry of the remnant
and to explore the different emission scenarios. A full sensitive X-ray
coverage of the SNR has not been achieved before; the ROSAT soft X-ray survey
has only revealed few photons from the source \citep{Tian08}.
So far, an \xmm pointing towards the  north-east and two \emph{Suzaku} pointings
towards the north-west and south-west of the remnant have been analysed in view
of the SNR \citep{Tian10,discovery_paper, bamba12}. The \xmm data have revealed
the absorption pattern mentioned above, which has been confirmed with
\emph{Suzaku}. No thermal emission has been found from the spectral analysis of
the data. The photon index of the power-law  model fitted to various subregions
of the remnant has been reported to vary between 2 and 3, so far without a
clearly identified morphological pattern. \cite{bamba12} have found one region
with a particularly hard index ($\sim1.8$); however, the angular resolution of
the telescope might have masked the contribution from strong point sources to
the spectrum.

For the work presented in this paper, we have obtained and analysed data from
\xmm covering the full SNR extent. For the first time, we provide a full
X-ray image of the remnant (see Fig.~\ref{fig:xmm_gallery}) as well as a
consistent spectral analysis across the source. The image has been divided into
a grid of eighteen source regions (with three background regions), and the corresponding
spectra have been used to analyse the spatial variations of the X-ray source
parameters across the remnant. The data permit  the X-ray map to be compared with
the sky maps in the TeV, radio, and submillimetre bands. The analysis of
the \xmm data has been hampered to some degree by the presence of significant
stray light coming from a nearby source (likely 1RXS J173157.7$-$335007) and Galactic
ridge emission. An extensively tuned analysis has therefore been performed to deal with that
problem, which is described below. The paper is organised  as follows: In Sect.
2 the \xmm data set and analysis is described in detail; Sect. 3 deals with
connection of the obtained results between  X-ray and other bands; and Sect. 4 summarises
the conclusions.

\section{Data analysis}
\subsection{Observations and data reduction} The remnant and the CCO have been
observed with \emph{\xmm} several times. The list of observations is presented
in Table~\ref{tab:obs}. For data screening and reduction the \emph{XMM~SAS}
version 20141104\_1833-14.0.0 was used (this version also includes 
the formerly independent extended source analysis software package ESAS). For
the analysis of extended emission from the remnant we followed the procedures
described in the
ESAS~Cookbook\footnote{http://heasarc.gsfc.nasa.gov/docs/xmm/esas/cookbook/xmm-e
 sas.html} unless stated otherwise. We excluded data taken in windowed and
timing modes, and only considered observations performed in imaging mode. We
also excluded data from CCD chips operating in the  `anomalous mode'
characterised by a high level of noise at low energies, and screened the data for
periods of enhanced orbital background (using the ESAS tasks
\emph{mos/pn-filter} with default filtering criteria).

Inspection of the images extracted from the screened data immediately revealed
that all observations are affected by stray light from a nearby bright source
(likely 1RXS J173157.7$-$335007) to some extent. The X-ray  stray light in the EPIC
cameras is mainly produced by single reflections of rays coming from outside of
the field of view from the hyperbolic section of the mirrors and is detected as
spurious arc-like structures across large fractions of the detector area.
Fortunately, the projection of these onto the sky plane changes from camera to
camera and also with pointing direction. Therefore, even after visual-based
exclusion of the regions contaminated by stray light, the remnant was still completely covered by observations, although the exposure was reduced, as can be
seen in the cumulative exposure map presented in Fig.~\ref{fig:ov}.

\begin{figure}[t!]
    \centering
        \includegraphics[width=\columnwidth]{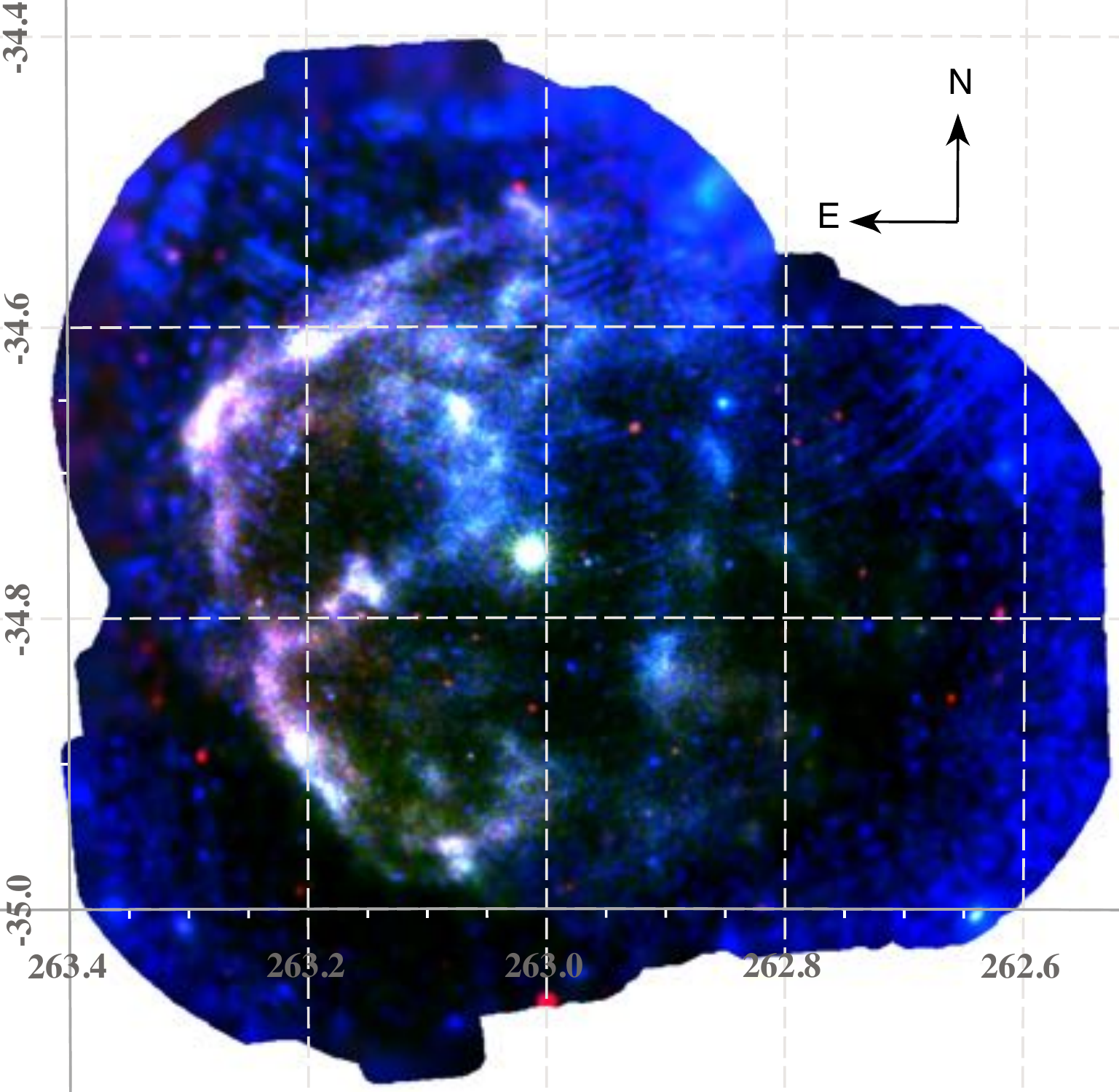}
        \includegraphics[width=\columnwidth]{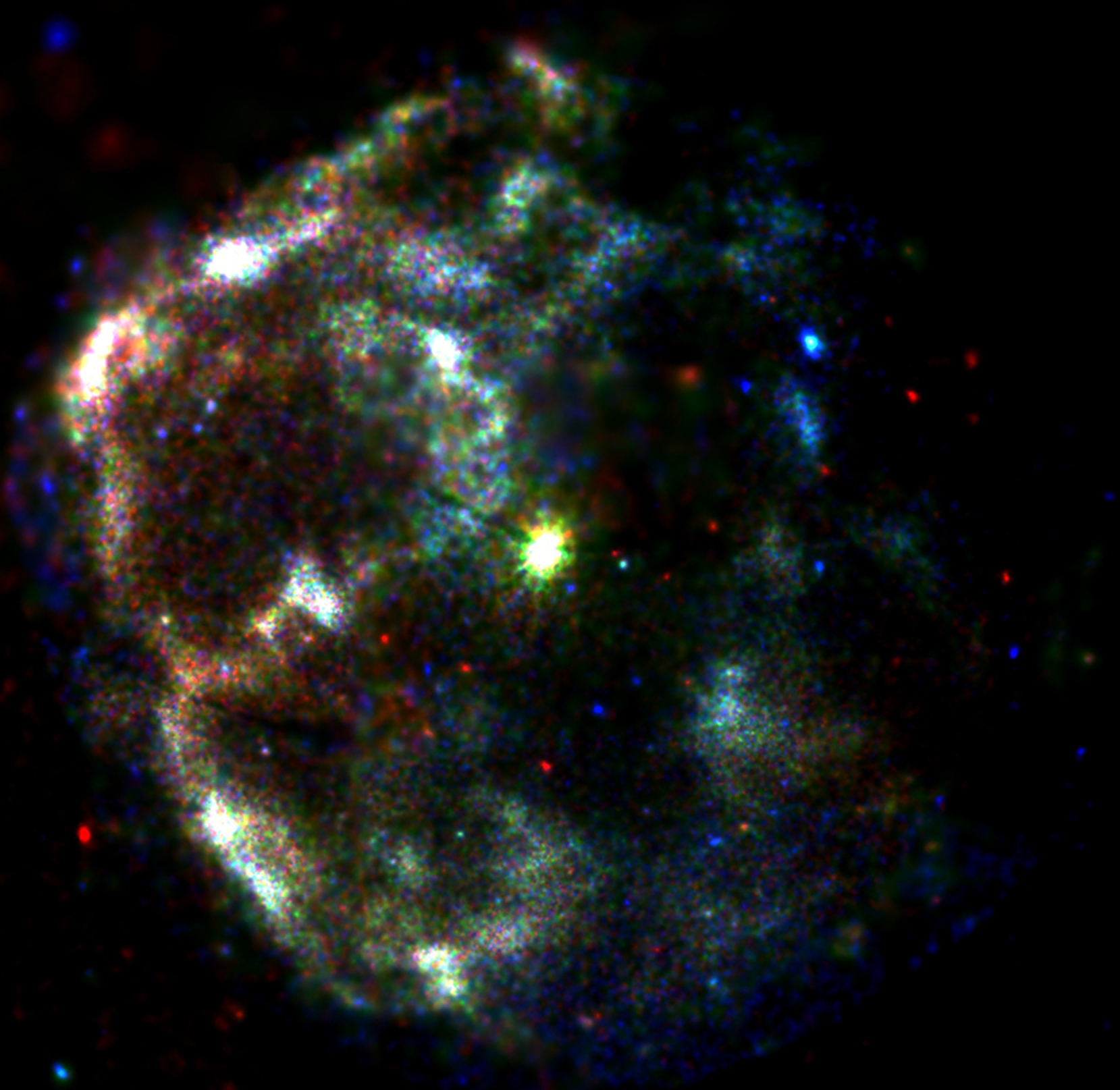}
    \caption{Pseudo-colour mosaic image of all observations showing 0.4-1.8 keV (red), 1.8-2.8 keV (green) and 2.8-10.0 keV (blue) 
    rates corrected only for
    the residual proton background (top panel), and central part of the same
    image corrected for all identified background components (bottom) panel.
    All images are presented in equatorial coordinates shown in a grid in the top image.
    The scale of uncorrected image has been adjusted to emphasise the
    contamination by stray light (mostly seen in blue) and extended emission towards the western part
    of the remnant. Note that the stray-light contamination is even more
    severe in individual observations as the location of the single
    reflection arcs is sensitive to the pointing of telescope.}
    \label{fig:xmm_gallery}
\end{figure}

Our study is focused on the extended emission of the remnant shell, so the
contribution of the point sources must be removed. To detect the point sources,
for each observation and camera (i.e. each individual exposure) we extracted
images and ran the SAS source detection pipeline (using both \emph{eboxdetect}
and \emph{emldetect} tasks) for four energy bands: 0.4-1.8, 1.8-2.8, 2.8-10,
and 0.4-10\,keV. In addition, we followed the same procedure for the combined
mosaic image including all observations. All detected sources were then merged
into a single list, which was then manually screened to exclude the spurious
detections corresponding to bright shell filaments. The final source list
consisting of 110 sources was then used to mask the point sources for both
imaging and spectral analysis (using the \emph{region} task to define
exclusion regions containing 95\% of flux for each source).

We also used the multi-colour mosaic presented to define source extraction
regions for spectral analysis. We considered both variation in total intensity
and colour  (i.e. source intensity in individual energy ranges) to define the
22 regions matching the observed structure of the shell and the
background in order to investigate spatial variations of the spectrum across the
remnant (see Fig.~\ref{fig:ov}).

For each observation and telescope we derived the spectral products (i.e.
source and in-orbit particle background spectra and respective response files)
for visible regions using the SAS tasks \emph{mos/pn-spectra}. In total, 167
spectra with a mean exposure of $\sim42$\,ks were derived. For each spectrum, the exposed sky
area (excluding areas contaminated by stray light and point sources, gaps between
individual CCD chips, and bad pixels) and scaling for the residual soft proton
contamination (see the next section) were calculated using the task
\emph{proton\_scale}.

\begin{table}[b]
        \begin{center}
        \begin{tabular}{cccc}
                Observation     & Observation  & Duration, & Exposure  \\
                     ID      &      Date     & ks        & (M1/M2/PN), ks\\
                \hline
                0405680201 & 2007-Mar-21 & 25.4 & 22.5/22.9/12.1\\
                0673930101 & 2012-Mar-02 & 23.7 & 21.9/21.8/0\\
                0694030101 & 2013-Mar-07 & 72.4 & 53.1/54.9/42.5\\
                0722090101 & 2013-Oct-05 & 61.3 & 53.3/54.1/41.6\\
                0722190201 & 2014-Feb-24 & 131.2 & 88.4/97.4/0\\
                total: & & 314 & 239.2/251.1/96.2 \\
                \hline
        \end{tabular}
        \end{center}
                \caption{List of \xmm observations of \src. The `Exposure' column reflects
                the effective imaging exposures for each camera (EPIC MOS1, MOS2, and PN). Zero
                exposure means that PN was operated in timing mode.}
        \label{tab:obs}
\end{table}

\subsection{Spatially resolved spectral analysis} 
A spectral analysis of extended source emission requires careful  modelling of the
background. We estimated the contribution of the in-orbit particle background
and out-of-time events (for EPIC PN) based on the count-rate in the unexposed
detector areas using the SAS tasks \emph{mos/pn\_back} and subtracted resulting
spectra from all observed spectra. To estimate the local
cosmic background, we used the source-free regions 0, 1, 2 (only available in
observations 0405680201, 0694030101, and 0722090101, respectively, i.e. those with
pointings displaced from the shell centre).

The source resides in the Galactic plane, so the cosmic background is expected
to be dominated by the Galactic ridge emission of unresolved point sources
(\emph{GRXE}, \citealp{Revnivtsev06}). As recently discussed by \cite{yuasa12},
the 2-50\,keV X-ray spectrum of the \emph{GRXE} can be represented with an absorbed
two-temperature collisional ionisation equilibrium plasma (CIE) model with
temperatures of $\sim1.5$\,keV and $\sim15$\,keV. We found that the observed
local background in our observations in the 0.4-10\,keV energy range is well
described with an absorbed single-temperature CIE model with $kT\sim0.7$\,keV
(see also Table~\ref{tab:fit_res}). The discrepancy in the derived temperature
value is not surprising taking into account the difference in energy range and
likely spatial variation of spectral properties of the ridge emission and, as
we verified, has little effect on the derived parameters of the shell spectrum.
We  also verified that the choice of the absorption model does not
affect other parameters and use \textsl{wabs} model \citep{wabsref} throughout
the paper.

The contribution from the local hot bubble (which is also expected to have
$kT\sim0.7$\,keV, but is not expected to be absorbed) and from background AGN
were not formally required by the fit. However, to avoid a possible bias for
other parameters, we still included the well-studied AGN background as a fixed
absorbed power law with an absorption column density linked to that in a given
region (which is comparable with the Galactic absorption column in the
direction of the source), a photon index of $\Gamma=1.46,$ and a scaled
normalisation of
$4.6\times10^{-7}$photons\,keV$^{-1}$cm$^{-2}$s$^{-1}$arcmin$^{-2}$ at 1\,keV
\citep{agn_spe}. The actual sky area from which individual spectra were
extracted differs between observations, so all model components were scaled to
unit area with a scaling constant calculated using the \emph{proton\_scale}
task.

We found also that despite the exclusion of the enhanced orbital background
periods, a residual soft-proton contamination remained noticeable in all
observations and, in fact, given the low surface brightness of the remnant, is
a major background component. Therefore, we included a power law (or a broken
power law if required) not folded through the instrumental response in the
model to account for it. The normalisation of this component is known to depend
on the position in the detector plane. However, the relative normalisation for
two arbitrary regions can be calculated using the task \emph{proton\_scale}
based on calibration observations. Therefore, we fit the parameters of this
component only for a single arbitrarily chosen sky region in each observation,
and scale it for all other regions using the scaling calculated with
\emph{proton\_scale}.

Emission from the remnant itself had been reported to be non-thermal with a
pure power-law spectrum, possibly with spatially varying photon index
\citep{bamba12}. To model the source emission, we therefore used a power law
with independent normalisation and photon index for individual sky regions.
Once the general shape of the background spectrum was established, we included
all sky regions and fitted the source and background components simultaneously
to improve counting statistics. As already mentioned, two observations
including the longest one were centred on the CCO and contain no source-free
regions to estimate the background (including the residual soft-proton
background). This is another reason why we fit all spectra from all
observations simultaneously. The parameters of the source and cosmic background
spectra were assumed to be linked to a given sky region, and parameters of the
soft proton background to a given camera and observation. For modelling we used
the \emph{sherpa}\footnote{http://cxc.harvard.edu/sherpa4.4/} spectral fitting
package from the \emph{Chandra} interactive analysis of observations (CIAO v
4.7) software.

Initially we found that the shell spectrum significantly hardened in the northern
part of the remnant. However, the same area is most heavily contaminated by 
stray light. Some excess emission is also evident throughout most of the image,
and in particular in the northern background region. It is clear, therefore,
that stray-light contamination is not limited to the apparent arc-like
structures (and their immediate neighbourhood) already excluded from the
analysis, but extends much further and might even be present in exposures with
no visible sign of stray-light contamination. A more conservative spatial
filtering of all possibly contaminated areas would imply dropping the entire
northern half or even two-thirds of the remnant from the analysis, so we opted
to treat the residual stray-light contamination as an additional background
component. 

\begin{figure*}[t]
    \includegraphics[width=0.3\textwidth]{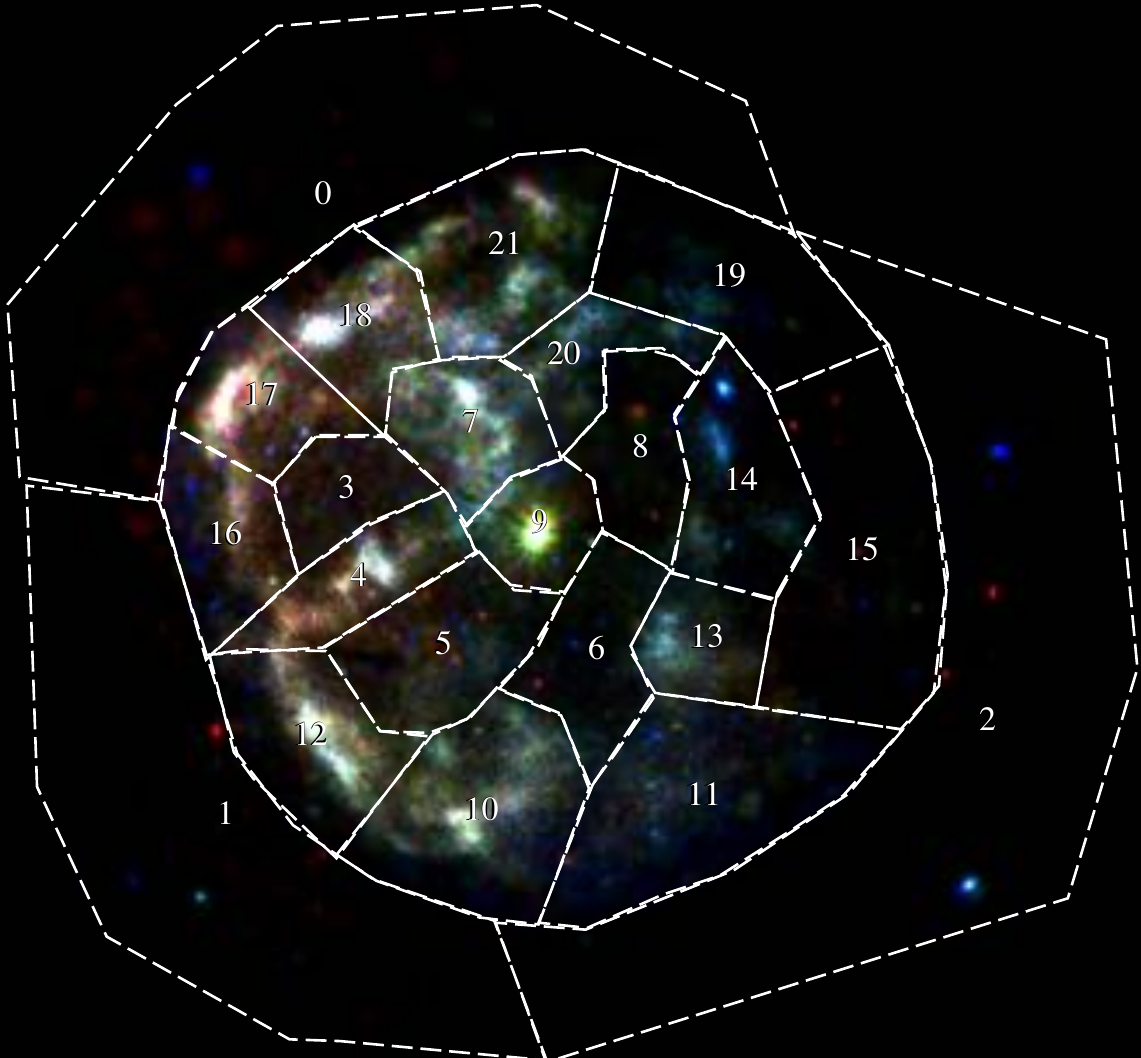}
\includegraphics[width=0.3\textwidth]{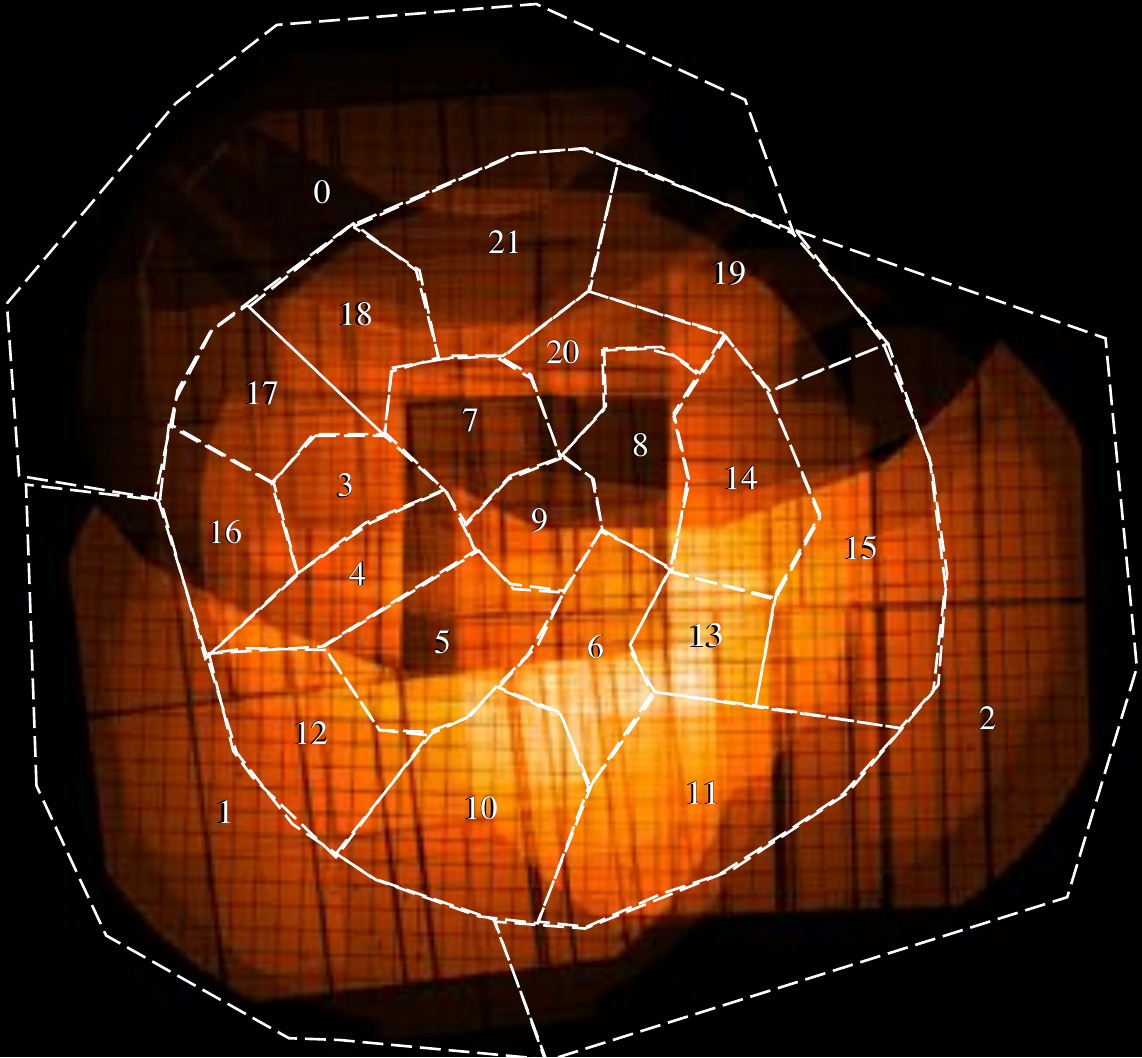}
\includegraphics[width=0.39\textwidth]{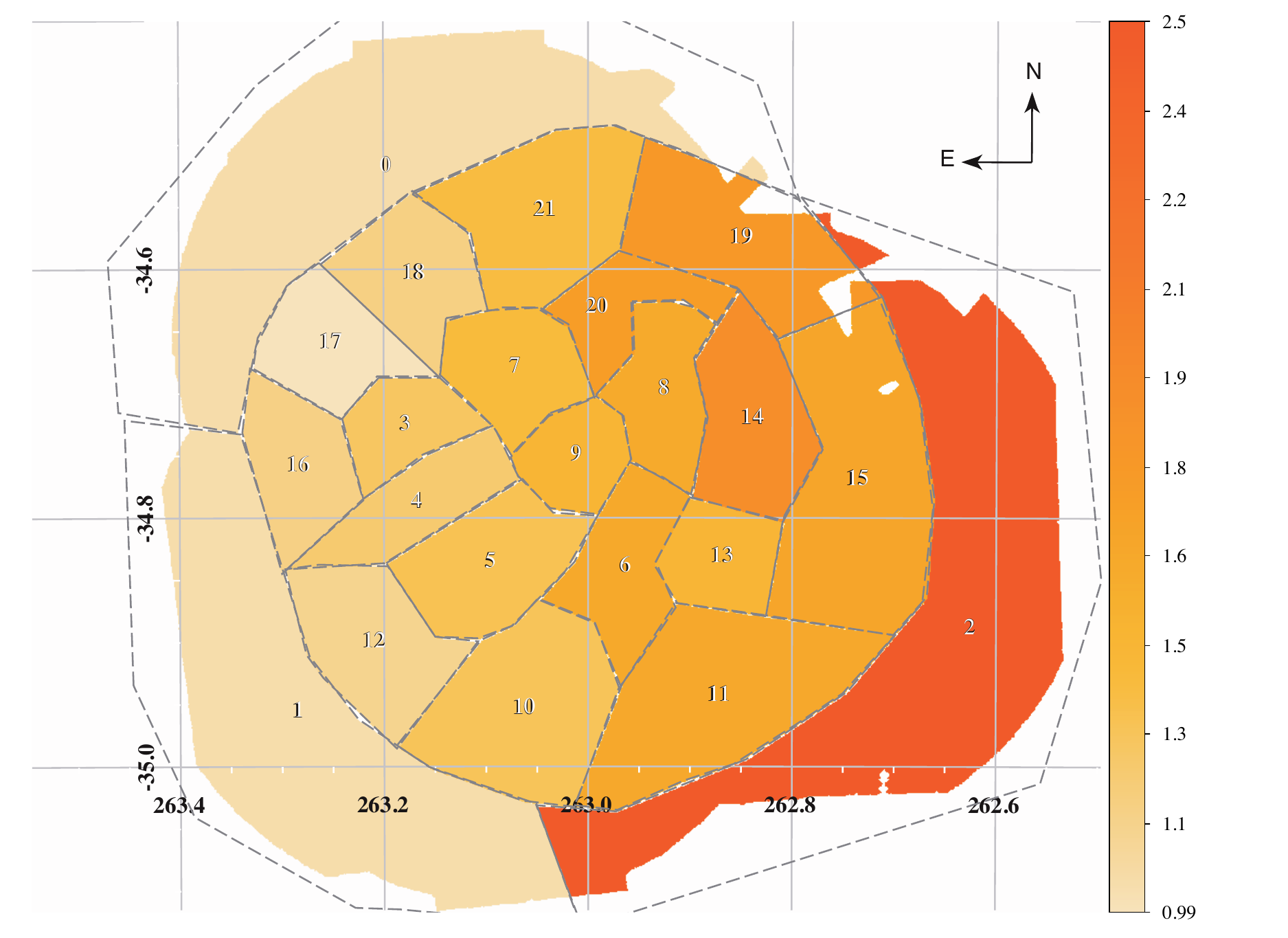}
\caption{Left panel: Mosaic pseudo-colour image in 0.4-1.8\,keV (red), 1.8-2.8\,keV (green), and 2.8-10\,keV (blue) energy bands corrected for particle, soft proton, and residual stray-light contamination. The definition of regions used for spectral analysis is also shown. Middle panel: Combined exposure map in 0.4-10\,keV band where regions most heavily contaminated by stray light and excluded from the analysis can be identified. 
Right panel: Colour-coded absorption column density.}
        \label{fig:ov}
\end{figure*}

\begin{figure*}[t]
    \centering
    \includegraphics[width=0.95\textwidth]{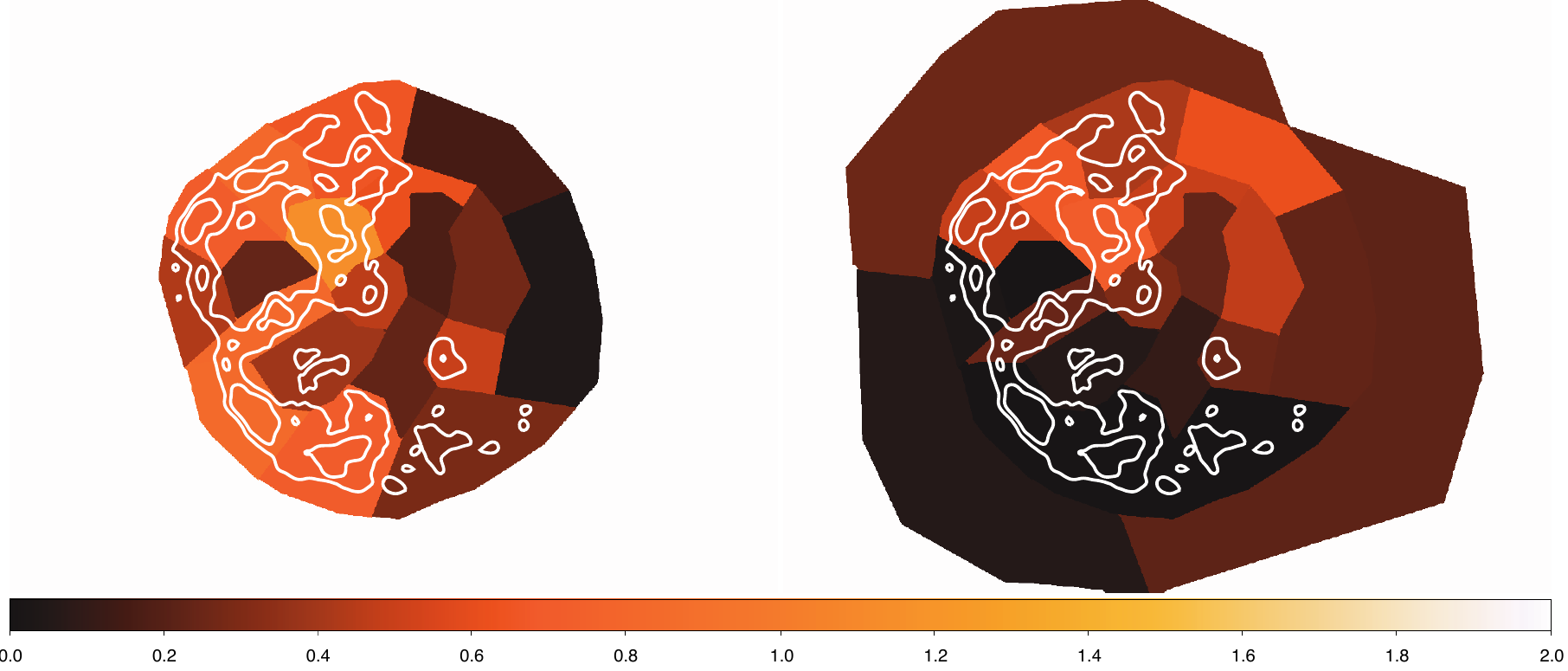}
\caption{Average surface brightness of the SNR emission (left) and residual stray-light contamination (right) in units
of $10^{-13}$erg\,cm$^{-2}$\,s$^{-1}$\,arcmin$^{-2}$ in the full (0.4-10\,keV) energy range. The stray-light emission spectrum 
is substantially harder, so it has a factor of two higher energy flux for a given photon flux. Surface brightness
roughly correlates with the reduction of exposure due to exclusion of the most heavily polluted regions 
shown in Fig.~\ref{fig:ov}. The contours show equal brightness levels for the extended
emission for reference.}
        \label{fig:stray}
\end{figure*}

The energy dependence of the effective area is not calibrated for stray light,
so it is not trivial to model this component. Furthermore, the degree of
contamination likely differs between individual observations for the same sky
region, which implies that for each spectrum an additional free parameter must be
constrained. We found, however, that a single power law with a photon index of
$\sim1.5$ and a normalisation only variable between individual sky regions but
not between individual spectra allows  all spectra to be fit. Unfortunately, the
normalisation of this component is also strongly correlated with the photon
index of the remnant spectra, so the spectral index of the extended emission
remains unconstrained for individual regions. Moreover, the remnant is located
close to the Galactic plane, so an additional contribution from the Galactic
ridge X-ray emission (GRXE) is expected. The spectrum of the ridge consists of
a number of emission lines and a power-law continuum with photon index of
$\sim2$, close to that of the remnant. The intensity and spectrum of the GRXE
are known to depend on the Galactic coordinates
\citep{Valinia98,Revnivtsev07,Ebisawa08}, and the vertical scale of the disc
component of the GRXE of $\sim1.5^\circ$ \citep{Revnivtsev07} is comparable to
the extension of the SNR. We attempted to include this component in the fit as
well. However, it was impossible to disentangle the contribution
of the three power-law continua with similar indices in individual regions.
Therefore, we had to conservatively assume constant photon indices across the
remnant for both the SNR and residual contamination from stray light/ridge
emission (referred to simply as stray-light contamination). The average surface
brightness of source and residual contamination emission are shown in
Fig.~\ref{fig:stray} on the same scale. We note that while the residual
contamination appears relatively strong, the SNR emission is significantly
softer than the stray-light emission, so the SNR emission dominates the photon flux
in all energy ranges where it is actually detected. The resulting fit is
statistically acceptable. However, it is still possible that we have overestimated
the stray-light contribution in some regions.

\begin{figure}[ht]
    \centering
        \includegraphics[width=\columnwidth]{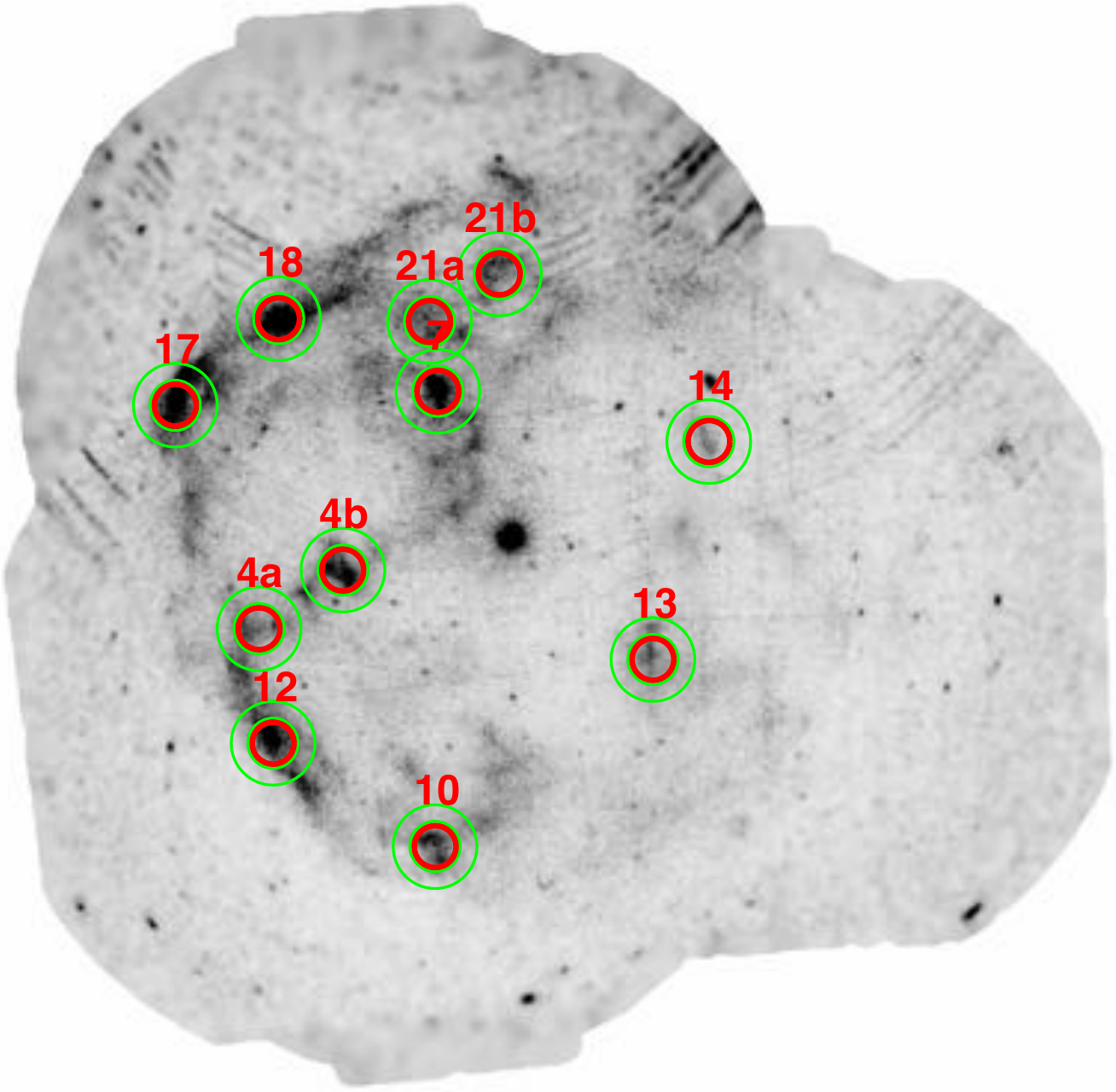}
        \includegraphics[width=\columnwidth]{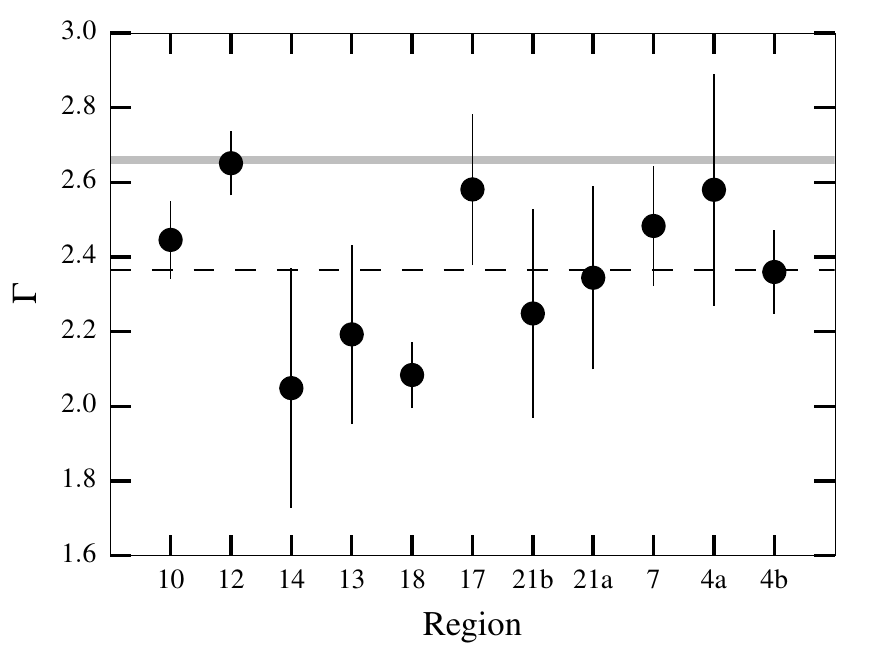}
    \caption{Definition of source and background regions used to probe
    variations of the photon index across the remnant (top, the image is in the 0.4-10\,keV band, adaptively smoothed 
    and corrected for exposure and quiescent particle background) and the
    best-fit values of the photon index for each region (bottom, $1\sigma$ uncertainties). The dashed line represents the mean value
    and the shaded area corresponds to the photon index estimated using entire shell. For consistency, the region names
    correspond to the larger regions introduced in Fig.~\ref{fig:ov}. }
    \label{fig:fil_anal}
\end{figure}

It is also important to emphasise that we  cannot exclude a spatial
variation of the SNR spectrum as it can be masked by changes in the
normalisation of the stray-light component. In order to verify that the photon index is
indeed stable throughout the shell we conducted a simplified but more
straightforward analysis for the brightest parts of the shell. In particular,
for several bright knots shown in Fig.~\ref{fig:fil_anal}, for each observation
and instrument we extracted spectra from small circular regions with radii of
$50^{\prime\prime}$, comparable with the \xmm point spread function (PSF), and associated backgrounds from
annuli centred on the same position with inner and outer radii of
$60^{\prime\prime}$ and $100^{\prime\prime}$, respectively. We visually
inspected individual images to exclude spectra where either source or
background regions were severely affected by stray light or were close to the
edge of the field of view. The remaining spectra were then modelled with an
absorbed power law after subtracting the background and grouping to 100 counts
per energy bin. We kept the absorption column free, as it is quite clear from
the optical extinction analysis presented in section~\ref{sec:dist_mc} that it
varies on scales as small as $\sim1^\prime$. The best-fit results are presented
in Fig.~\ref{fig:fil_anal}. While there is some scatter in the measured photon
index values, they seem to be  consistent with each other within the
uncertainties. The distribution of measured photon indices is also compatible
with that reported based on \emph{Suzaku} data \citep{bamba12}, as follows from
the high two-tailed p-value of 0.81 for the two-sample Kolmogorov-Smirnov
test. On the other hand, the average value is somewhat lower than that measured
using the entire shell (although the deviation is not significant). There are several
reasons for this offset, for example  a variation in the absorption column
within larger regions used above, the presence of unabsorbed components in
the background (i.e. the local bubble)  unaccounted for in the simplified filament
analysis, an overestimation of the stray-light contribution for the analysis of the
entire shell, or a combination of the above. It is important to note, however,
that the obtained results still seem to be robust despite the apparent
complexity of the analysis as justified by the clear correlation of the deduced
absorption column with the optical extinction described in
section~\ref{sec:conh}. We also do not expect these factors to significantly
affect the flux estimates, so this issue does not affect any of our conclusions. To
resolve it completely, additional observations with \emph{Chandra} which seem to
be less susceptible to stray light will, however, be required.

\begin{figure}[ht]
    \centering
        \includegraphics[width=\columnwidth]{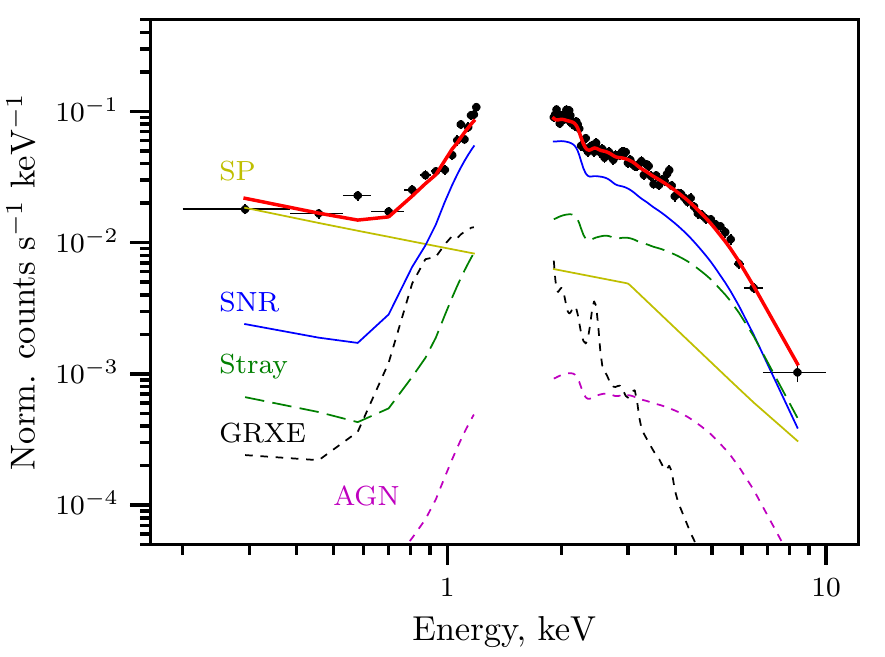}
    \caption{Representative QPB subtracted spectrum extracted from region 21 using
    the MOS2 data from observation 0722190201 (black points). Identified source and
    residual background components described in the text are also shown as labelled in the figure.}
    \label{fig:spe_ex}
\end{figure}

The final fit results are presented in Table~\ref{tab:fit_res}. All
uncertainties are at the $1\sigma$ confidence interval unless stated otherwise. Flux
uncertainties are estimated from a sample of $2\times10^4$ simulated spectra
with parameters distributed according to the best-fit results, i.e. they are not
affected by absorption. An example of a representative spectrum from region 21
which contains both significant source and background contributions modelled as described
above is presented in Fig.~\ref{fig:spe_ex}.

The spatially resolved spectral analysis presented above is also
essential for correcting the mosaic image presented in Fig.~\ref{fig:ov} for
the residual quiescent particle background (QPB), residual soft proton, and stray-light
contamination. Using the obtained best-fit parameters we created the quiescent
particle background and residual soft proton contamination images for each
exposure using the \emph{mos/pn-spectra} and \emph{proton} tasks. The
resulting images were then merged using the task \emph{merge\_comp\_xmm} to obtain mosaics
covering the entire remnant. To account  for the residual stray-light
contribution, we followed a similar approach generating mosaic images based on the
expected number of contaminating photons. In
particular, the average stray-light count rate in each region and energy range was calculated
based on the best-fit parameters presented above, and then multiplied by the
observed exposure maps to estimate the expected number of stray-light photons
in each sky pixel. The resulting count images for the quiescent particle, soft
proton, and stray-light backgrounds were then co-added and subtracted from the
combined net image. The final image was corrected for uneven exposure and
smoothed. The absorption-corrected image presented in
Fig.~\ref{fig:ext} was smoothed using a Gaussian kernel with constant width of
0.3$^\prime$. The same image without the correction for absorption
presented in Fig.~\ref{fig:ov} and Fig.~\ref{fig:xmm_gallery} was adaptively
smoothed using the task \emph{adapt\_merge} to contain at least 300 counts per
pixel in each band.

\begin{table}
        \begin{center}
        \begin{tabular}{p{1cm}cccc}
\multicolumn{5}{c}{Parameters common to all regions}\\
\hline
                        \multicolumn{3}{l}{$\Gamma_{SNR}$} &    \multicolumn{2}{c}{2.66(1)}\\
                     \multicolumn{3}{l}{$F_{SNR,2-10}$} &    \multicolumn{2}{c}{$3.77(7)\times 10^{-11}$\escm}\\
                      \multicolumn{3}{l}{$\Gamma_{Stray}$} &    \multicolumn{2}{c}{1.54(1)}\\
                     \multicolumn{3}{l}{$kT_{CIE, Ridge}$} &    \multicolumn{2}{c}{0.74(1)}\\
                                         \multicolumn{3}{l}{$\chi^2_{red}/dof$} &    \multicolumn{2}{c}{1.03/20227}\\

\multicolumn{5}{c}{Region specific parameters}\\
\hline
                                 Reg. & Area$^{a}$ &                  nH$^{b}$, & $F_{SNR, 2-10}^{c}$ &                               $F_{stray, 2-10}^{c}$\\

\hline
                                    0 &      272.3 &                    1.11(3) &                   - &                                              6.9(4)\\
                                    1 &      233.1 &                    1.06(2) &                   - &                                             1.25(8)\\
                                    2 &      326.1 &                     3.0(2) &                   - &                                              7.1(5)\\
                                    3 &       24.3 &                    1.16(2) &              0.57(4) &                                           0.0020(2)\\
                                    4 &       31.6 &                    1.21(1) &              2.6(1) &                                             0.76(5)\\
                                    5 &       44.2 &                    1.29(2) &              1.7(1) &                                             0.25(2)\\
                                    6 &       36.1 &                     1.6(3) &             0.83(7) &                                             0.37(2)\\
                                    7 &       33.1 &                    1.44(2) &              3.8(2) &                                              2.4(2)\\
                                    8 &       32.1 &                    1.68(6) &             0.50(0) &                                             0.75(5)\\
                                    9 &       22.3 &                    1.44(2) &              1.0(1) &                                             0.67(4)\\
                                   10 &       66.1 &                   1.297(1) &              4.7(2) &                                           0.005(4)\\
                                   11 &       80.5 &                    1.53(2) &              2.3(1) &                                           0.007(5)\\
                                   12 &       44.7 &                    1.14(1) &              3.7(1) &                                           0.004(3)\\
                                   13 &       25.3 &                    1.49(2) &             1.24(8) &                                             0.63(4)\\
                                   14 &       44.9 &                    1.93(4) &              1.2(1) &                                              2.1(1)\\
                                   15 &       94.2 &                    1.59(5) &             0.30(2) &                                              2.1(1)\\
                                   16 &       33.9 &                     1.1(2) &              1.5(1) &                                            0.065(4)\\
                                   17 &       33.3 &                    1.02(1) &              2.4(2) &                                              1.6(1)\\
                                   18 &       36.4 &                    1.21(2) &              3.0(2) &                                              2.4(2)\\
                                   19 &       61.0 &                    1.95(9) &             0.90(3) &                                              3.8(2)\\
                                   20 &       25.1 &                    1.81(5) &              1.5(1) &                                             1.22(8)\\
                                   21 &       60.1 &                    1.41(2) &              3.9(3) &                                              2.5(2)\\
        \end{tabular}
        \end{center}
        \caption{Spectral extraction region parameters and best-fit results assuming a pure power law with constant index for all regions.
        $^a$ in units of arcmin$^2$, $^b$ in units of $10^{22}$\,atoms\,cm$^{-2}$, $^c$ in units of $10^{-12}$\,erg\,cm$^{-2}$\,s$^{-1}$}
        \label{tab:fit_res}
\end{table}

\section{Discussion}

\subsection{Spectral energy distribution}
The broadband spectral energy distribution (SED) of the SNR has been discussed
by several authors \citep{discovery_paper,Yang14,Acero15} based on integrated
TeV and radio fluxes and X-ray flux detected from the eastern half of the
remnant scaled by factor of two. The latter assumption implies that
the X-ray flux was significantly overestimated as \xmm observations have revealed
that the eastern part is, in fact, much dimmer. The source is not detected in the GeV
band which could be useful to discriminate between leptonic and hadronic TeV
emission. \cite{Acero15} obtained stringent upper limits from Fermi
data which were interpreted as favouring a leptonic scenario. Here we present an updated
spectral energy distribution fit using the X-ray flux integrated over the entire
remnant, and assuming a simple single zone synchrotron-inverse Compton model.
It is important to emphasise that this is probably an
oversimplification as both radio and X-ray emission clearly show variation of
intensity/power-law slope \citep{Nayana17} over the remnant. However, the quality of
radio, X-ray, and TeV data does not permit a more detailed analysis at
this stage, so we stick to the model used in previous investigations to facilitate the
comparison of the results. Fitting and modelling of the SED were carried
out using the \emph{naima} package \citep{naima}.

\begin{figure}[h!]
    \centering
        \includegraphics[width=\columnwidth]{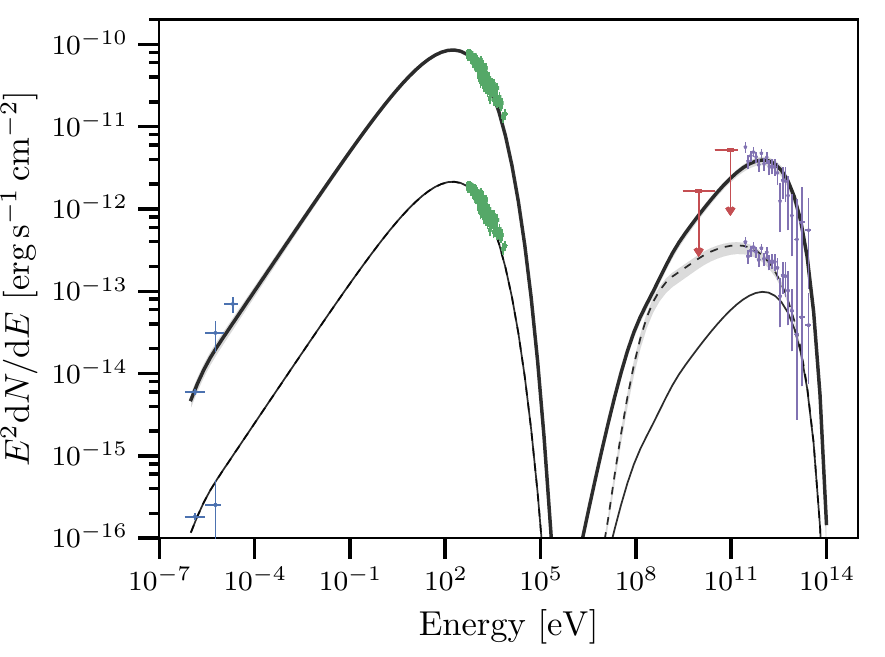}
    \caption{Broadband spectral energy distribution (SED) and the best-fit leptonic scenario model (upper set of lines).
    The X-ray spectrum used here is a simulated \emph{PN} spectrum with an exposure
    adjusted to yield the same spectral index and flux uncertainties as the best-fit result presented in Table~\ref{tab:fit_res}
    (i.e. the flux is integrated over all extraction regions with the contribution of point sources removed).
    The lower set of lines shows the SED scaled to represent fluxes within region 19 and the
    corresponding leptonic model with all parameters fixed to that from a global fit. The same model with
    an additional hadronic component accounting for the excess TeV flux is also shown for illustration (dashed line).}

    \label{fig:sed}
\end{figure}

In addition to the updated X-ray flux estimate,  we include the new
measurement at 325\,Mhz by \cite{Nayana17} and the radio fluxes at 1.4 and 5~Ghz reported
by \cite{Tian08}. We also included an additional seed photon component associated with the optical star located close to the
centre of the remnant with a temperature of $9\times10^4$\,K and a density of
$0.02$\,eV\,${\rm cm}^{-3}$ \citep{Doroshenko16},
which was not included in fits by \cite{Acero15}. The best fit presented in
Fig.~\ref{fig:sed} yields values similar to those reported by \cite{Yang14} with
an electron population index of $s_e=1.85(1)$ and a cut-off at 11.2\,TeV, background
field of $B=23(1)$\,$\mu$G, and an electron population energy above 1\,GeV of
$W_e=0.20(1)\times10^{48}$\,erg (again assuming a distance of 3.2\,kpc).

The derived magnetic field is consistent with earlier estimates and seems   to
be rather low compared with other young SNRs. Indeed, fields
derived from thin X-ray filament structures or the time variation of
X-ray emission were found to be significantly higher in some cases \citep[a few 100
$\mu$G;][]{2003ApJ...589..827B,2005ApJ...621..793B,2007Natur.449..576U}.
On the other hand, the filaments in HESS~J1731$-$347 do not seem to be  as sharp as in those cases, and rather clumpy,
i.e. they are more similar to those in RCW~86 \citep[whose age is
$\sim$2000~yrs;][]{2005ApJ...621..793B}.
The magnetic field in this SNR can thus indeed be lower than in other younger SNRs,
which would also support a leptonic origin for the bright VHE gamma-rays.

It is important to note that the single-zone SED model considered above is almost
certainly an oversimplification. Indeed, as discussed by \cite{Nayana17} and
below in section \ref{sec:morpho}, the morphologies of radio, X-ray, and TeV
shells are significantly different. Furthermore, based on the comparison
of 325~and 610\,Mhz fluxes, \cite{Nayana17} reported slightly different
spectral indices for the eastern and western parts of the radio shell. In
particular, the spectrum of the brighter eastern part was found to be somewhat
steeper, which, together with the spatial anticorrelation of radio and TeV
brightness, was interpreted as evidence for a stronger ambient magnetic field in
this region.

We also considered, therefore,  modelling of the SED of the north-western
part of the shell only, where the difference in radio/X-ray and TeV
morphologies is most apparent. It is important to emphasise, however, that the
quality of the data in all bands is not sufficient for a detailed spatially
resolved analysis of the SED and the estimates below are thus very approximate and
serve mainly an illustrative purpose.

For the estimates we consider region 19 defined for our X-ray analysis,
which roughly corresponds to the radio filament 3 reported by \cite{Nayana17}.
We estimate that X-ray, radio, and TeV fluxes from this part of the shell
roughly correspond to $\sim3, 2.5$, and 7\% of the total source flux. To
estimate the TeV flux we used the ratio of excess counts from region 19 to that
from the entire remnant using the image from \cite{discovery_paper}, and scaled the
spectrum reported there using this value. Similarly, we re-scale the X-ray
spectrum used for the global SED analysis to flux measured from region 19 as a
rough estimate. The resulting SED is presented in Fig.~\ref{fig:sed} along with
the global SED and a leptonic model with all parameters, except the
normalisation fixed to global values.

As can be seen from the figure, the model in this case significantly
underpredicts the TeV flux (or overpredicts the flux in the lower energy bands). There
are several possibilities to explain this discrepancy. For instance, as
suggested by \cite{Nayana17}, one can assume an ambient
magnetic field that is lower by a factor of two  in this region ($\sim13$\,$\mu$G) to reduce the relative flux at low
energies. Alternatively, the TeV excess can be attributed to a higher seed photon
density associated with the infrared emission from dust in the Galactic plane
or to an additional hadronic component. 
In the latter case, even arbitrarily assuming the same normalisation for proton and electron
populations, one can estimate density of the target medium required to account for the discrepancy at
$\sim2000$\,cm$^{-3}$. We note that this is still reasonably low as the X-ray absorption is highest in this direction
($\sim2\times10^{22}$\,atoms\,cm$^{-2}$), and even higher density might be required
to produce the observed CO emission \citep{Maxted15} (see also discussion in
section~\ref{sec:morpho}).The corresponding SED model is also shown in
Fig.~\ref{fig:sed} for reference.

\subsection{correlation of $^{12}CO(J=1-0)$  and X-ray absorption}
\label{sec:conh}
\cite{discovery_paper} estimated the distance to the source 
comparing the largest measured X-ray absorption column with the column density derived based on the velocity
spectra of $^{12}CO$ \citep{Dame01} and HI \citep{Haverkorn06} emission.
They concluded that these become comparable for velocities integrated
up to a radial velocity relative to the {local standard of rest} (LSR) of
-25\,km\,s$^{-1}$. The corresponding peak in
integrated column is located at -18\,km\,s$^{-1}$, which corresponds to a lower limit on the distance
of $\sim3.2$\,kpc. This estimate is qualitative, and in principle
depends on the assumed composition of ISM and needs  verification.

\begin{figure}[t!]
        \centering
                \includegraphics[width=0.49\textwidth]{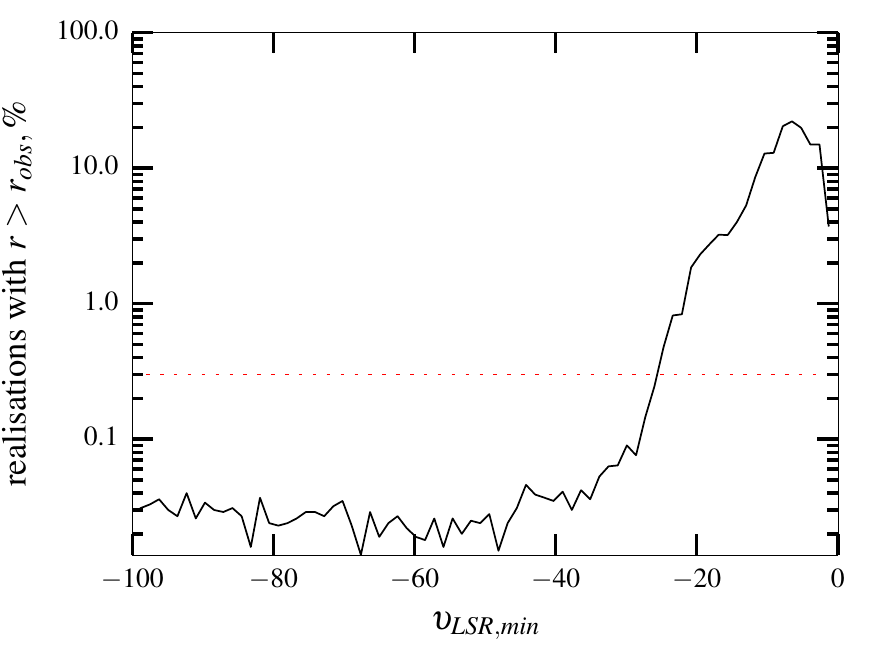}
                \caption{Significance of the correlation between the absorption column
                density derived from \xmm X-ray observations and the integrated CO temperature
                for different regions based on a permutation test (red dotted line indicates the $3\sigma$ level). The rapid drop in the
                correlation significance for $\upsilon_{LSR}\gtrsim-25$\,km\,s$^{-1}$ corresponds
                to a lower limit on distance of $\sim3$\,kpc.}
        \label{fig:co_nh}
\end{figure}

\cite{discovery_paper} pointed out that for the same integration velocity the
observed CO intensity increases in the direction of the Galactic plane
similarly to the X-ray absorption. This conclusion can be verified quantitatively
using the \xmm spatially resolved spectroscopy. To do so we use the CfA
survey \citep{Dame01} as used by \cite{discovery_paper}, and calculate the
Pearson correlation coefficient between the observed X-ray absorption and the
$^{12}CO$ column integrated up to a given velocity within a given spatial region.
None of the published HI data cubes \citep{Haverkorn06} covers the entire remnant
which complicates a similar analysis for this data set, due to the presence
of edge artefacts in individual data cubes covering parts of the remnant.
On the other hand, for a correlation analysis no additional assumptions on the ISM composition are required
as we do not have to convert the observed $^{12}CO$ line intensity to the
absorption column density and only rely on a spatial correlation of the two
quantities. To estimate the significance of the correlation in each case we
compare the observed correlation coefficient with a respective sample of values
obtained by shuffling the regions (i.e. a bootstrapping technique with $10^5$
realisations). As illustrated in Fig.~\ref{fig:co_nh}, the correlation is
indeed significant for integration velocities
$\upsilon_{LSR}\lesssim-20$\,km\,s$^{-1}$. We note that our result
is based on $^{12}CO$ data which exhibits no significant features around this velocity, so the
correlation is likely disrupted by an additional contribution from atomic hydrogen
which appears around $-18$\,km\,s$^{-1}$ \citep{discovery_paper}.
This confirms the suggested similarity of X-ray absorption and column density maps
\citep{discovery_paper} and reinforces their conclusions on the lower distance
limit although this analysis would certainly benefit from the detailed HI/$^{12}CO$ mapping of the entire remnant.

\begin{figure}[t!]
        \centering
                \includegraphics[width=0.49\textwidth]{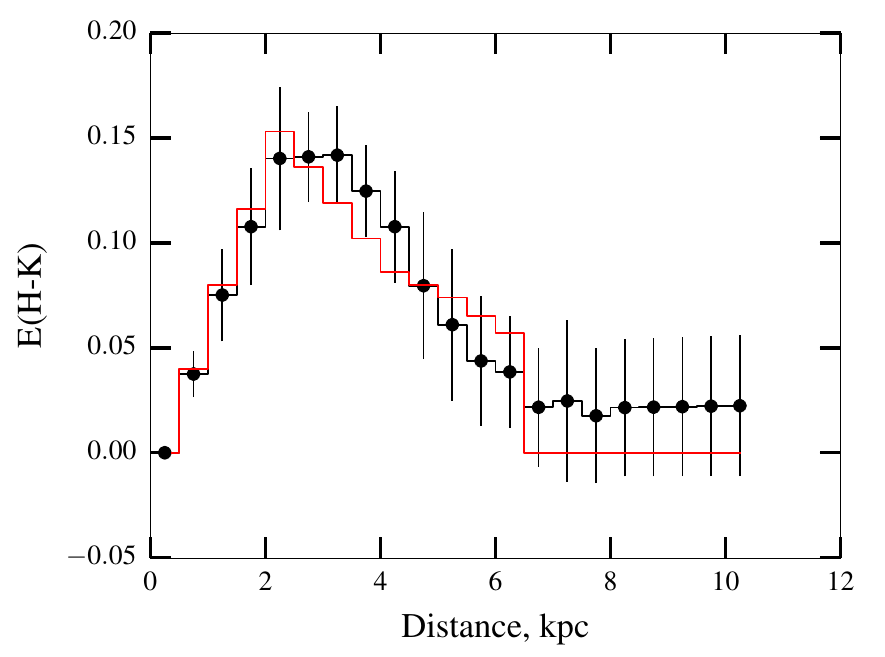}
                \caption{Mid-infrared extinction as a function of distance in the direction of the remnant. 
        The points and uncertainties correspond to mean and standard deviation of values reported by \cite{Schultheis14}
        averaged within a circle with radius of 0.25$^\circ$ centred on the CCO. The red line shows the extinction as a function
        of distance for the dense filament marked in Fig.~\ref{fig:extinction} (the respective
        uncertainties are not shown for clarity, but they are a factor of two smaller than for the black points).}
        \label{fig:3dext}
\end{figure}

It is interesting to note that a similar distance estimate can be obtained
independently based on the analysis of the optical extinction known to be
correlated with absorption in X-rays \citep{Vuong03}. Using the 3-D  dust
extinction mapping results presented by \cite{Schultheis14}, we estimated the
average absorption as a function of distance in the direction of the remnant as
shown in Fig.~\ref{fig:3dext}. Most of the absorbing material is concentrated
at around 3\,kpc, i.e. it coincides with the $^{12}CO$ emission peak at
$\upsilon_{LSR}\sim-18$\,km\,s$^{-1}$. Unlike the radio data, this
result does not depend on the Galactic rotation curve model. Finally, a distance
of about 3.4~kpc for the remnant is also suggested from the red-clump method
\citep{Zhu15} (priv. comm.).

\subsection{TeV, radio, and X-ray morphology of the shell}
\label{sec:morpho}
With the full remnant now covered also by X-ray observations, it becomes possible
to compare the morphology of the shell in TeV, radio, and X-ray bands. The TeV and
radio shells have been reported to have relatively flat azimuthal profiles
\citep{discovery_paper}, which does not seem to be the case for X-rays
where the  western part of the remnant is dimmer than the rest of the shell.

\begin{figure*}[t!]
    \includegraphics[width=0.33\textwidth]{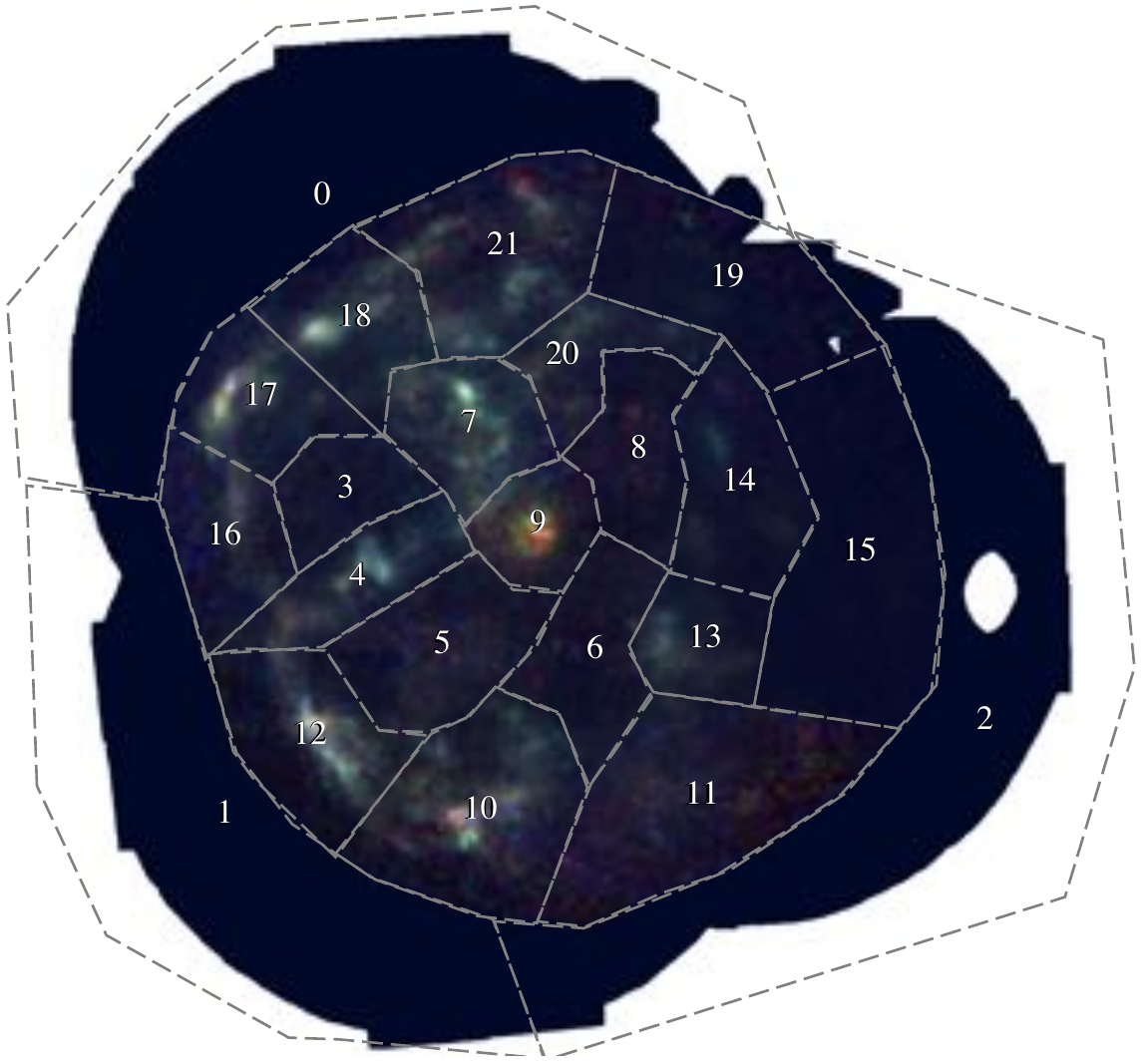}
        \includegraphics[width=0.33\textwidth]{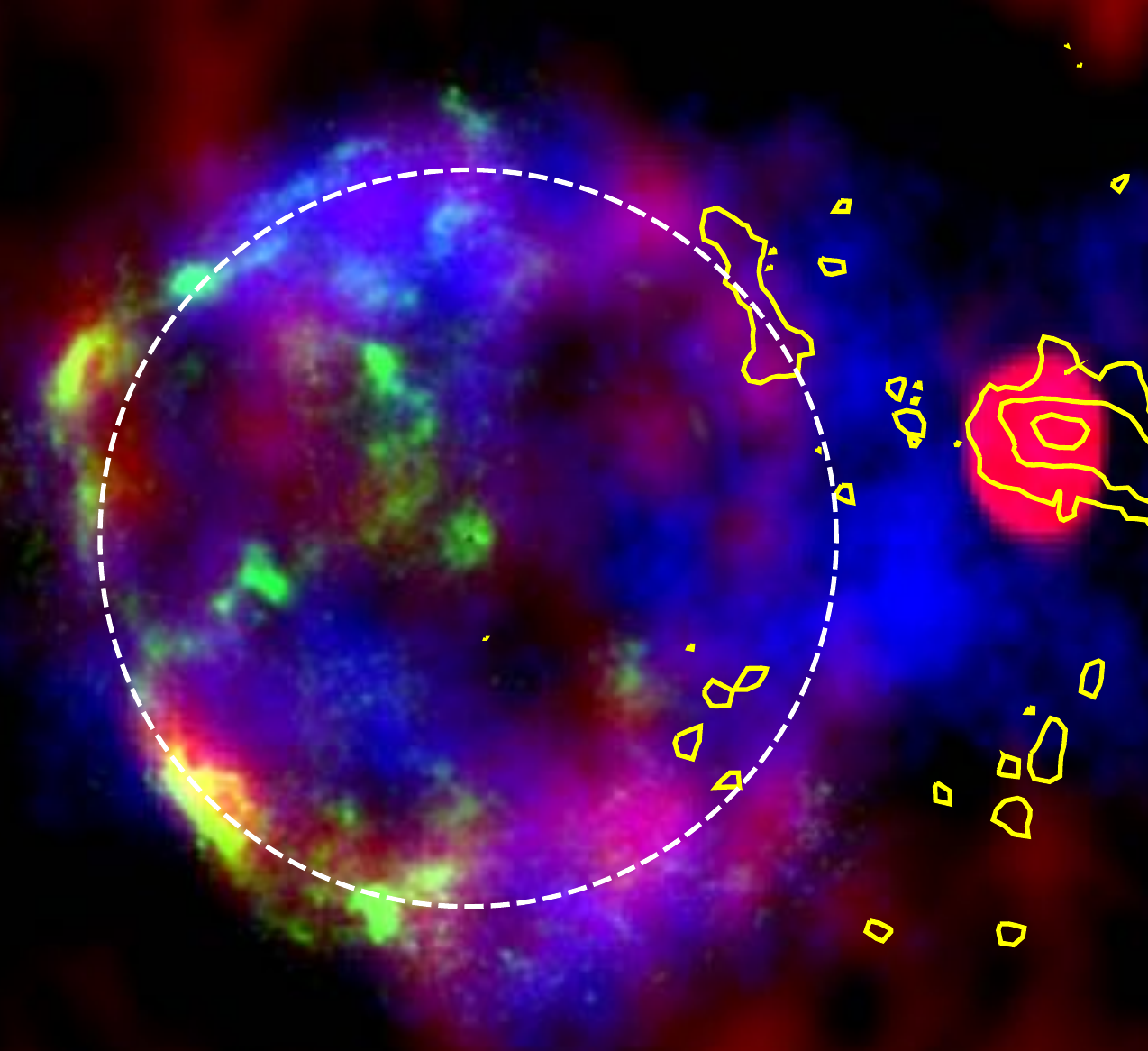}
        \includegraphics[width=0.33\textwidth]{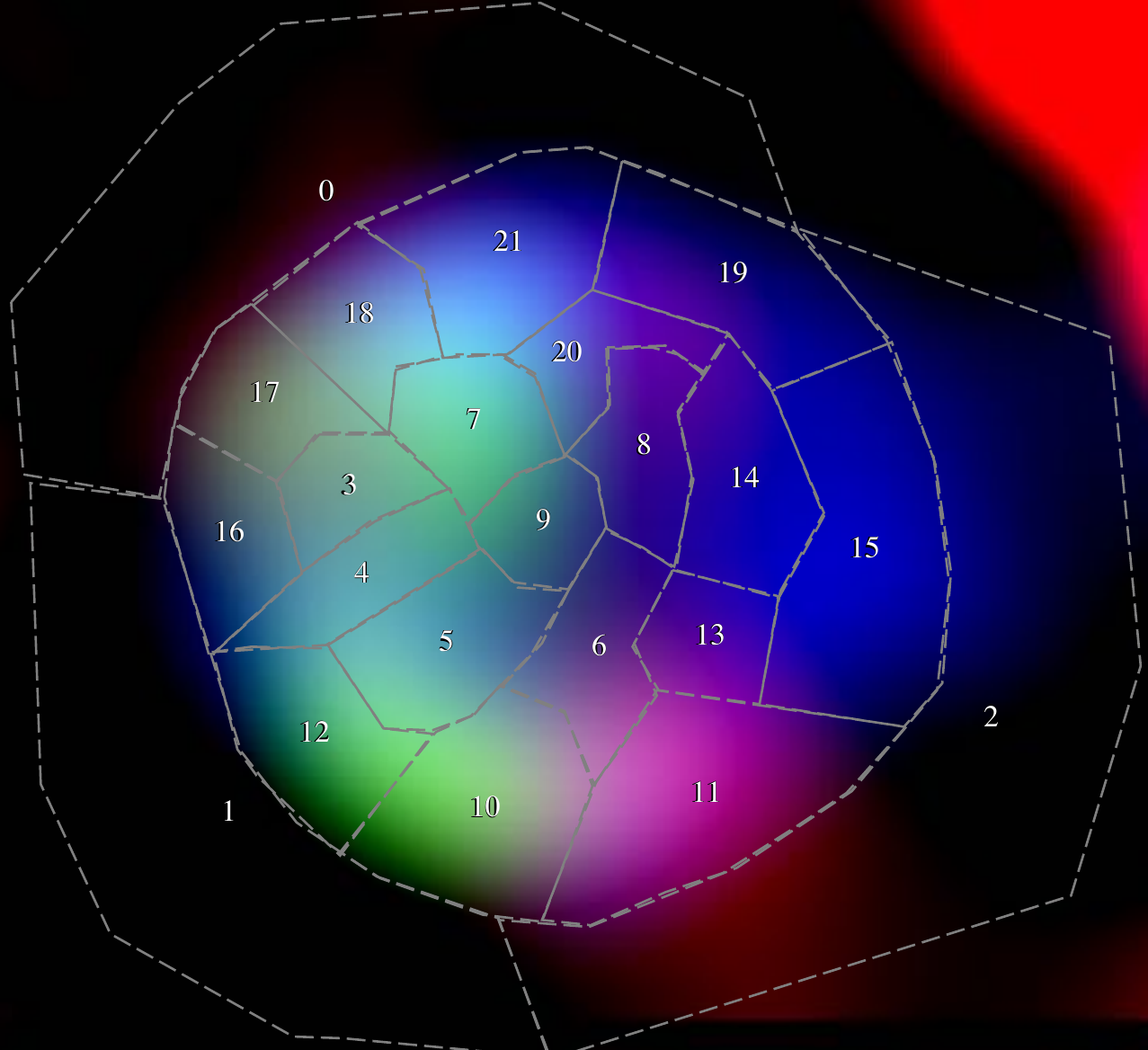}
    \caption{ X-ray image screened for point source contribution and corrected
for the identified background components and absorption representing
the intrinsic source flux (left panel). The bright blob in the centre of
the X-ray shell is due to residual emission from the bright CCO. A comparison of the SNR shell morphology in radio (red, GMRT), X-ray
(green, \xmm), and TeV (blue, HESS) bands is shown in the middle and right panels. The dashed circle
centred on the CCO indicates an approximate extension of the SNR shell based on the X-ray and radio
intensity. The
bright radio and X-ray filaments seem to be well correlated (middle panel).
However, the radio shell is more symmetric and extends further towards the
Galactic plane. There seems to be a correspondence between the continuum radio
and sub-mm emission (yellow contour) tracing the cold dense region also noted by \cite{Maxted15}
in the CS MOPRA data. The right panel shows all three bands smoothed with
a 0.06$^\circ$ Gaussian, corresponding to the PSF of the HESS data. We note  region 15 at
$l\sim353.5$, $b\sim-0.5$ where the shell is only bright in TeV.}

    \label{fig:ext}
\end{figure*}

For a meaningful comparison, however, it is important to account for non-uniform
absorption. The apparent blue tint of the false-colour mosaic image
(Fig.~\ref{fig:xmm_gallery}) is mostly due to the absorption of the soft flux by
the interstellar medium and not to a significant change in hardness of the SNR
emission, so the absorption is important. However, the overall impact of the
absorption on the observed X-ray flux is  minor as the shell spectrum is
relatively hard and the absorption only varies by a factor of three.
To better illustrate this point and enable a direct comparison of the X-ray shell
morphology with other bands, we estimated the intrinsic X-ray flux based on the
results of the spectral analysis. The corrected image reveals no significant
variation in hardness of the extended emission as illustrated in
the left panel of Fig.~\ref{fig:ext}. The overall intensity of the X-ray flux and the shape of the
shell are also not significantly affected, which is actually not surprising
given the relatively hard spectrum of the shell and  that the absorption
column density only varies by factor of three across the remnant. We use, therefore,
the full-band (0.4-10 keV) X-ray image for comparison with the radio and TeV maps.

\begin{figure}[ht!]
    \centering
        \includegraphics[width=\columnwidth]{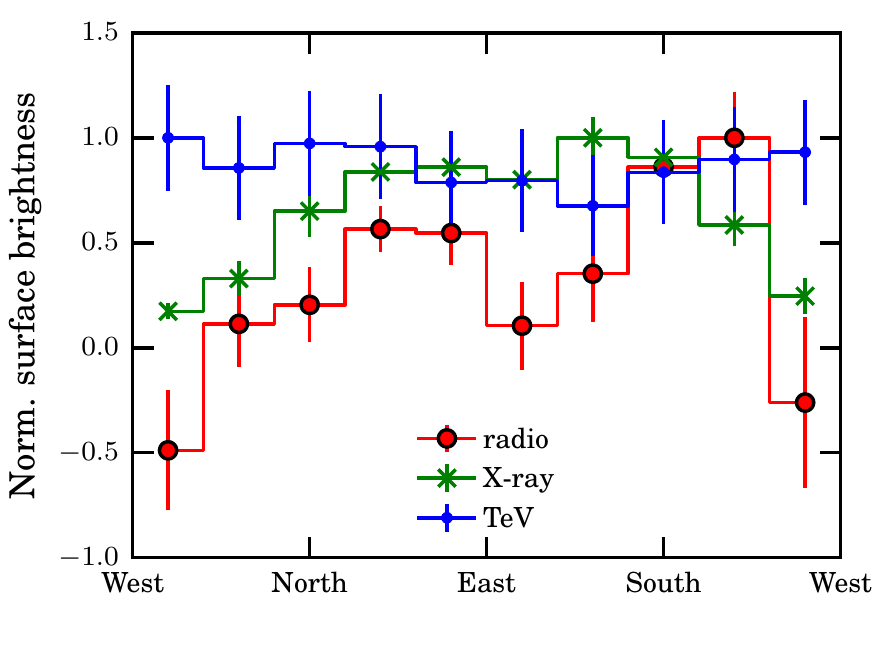}
    \caption{Shell azimuthal surface brightness profile in TeV, X-ray, and radio bands. The shell is assumed to be centred on the CCO and have radius from 0.18 to 0.24 degree. The radio and X-ray images were smoothed with HESS psf to allow meaningful comparison.  X-ray and radio emission are both suppressed towards the Galactic plane (i.e. western or spectral extraction regions 15,19), whereas the TeV profile is consistent with being flat.}
    \label{fig:azprof}
\end{figure}

The morphology in the radio band seems to depend on frequency
\citep{Nayana17}. However, a detailed analysis of the radio data is still ongoing, so
we limit the comparison to the 325~MHz band where the shell seems to be best
defined. For a quantitative comparison we also smooth all three maps with a
3.6$^\prime$ Gaussian kernel corresponding to the point spread function of the TeV map.
We also excluded the bright HII region to the west of the remnant prior to
convolution of the radio map, as illustrated in Fig.~\ref{fig:ext}. Using the
convolved maps, we obtained azimuthal surface brightness profiles for all three bands similar to those
reported by \cite{discovery_paper}. Taking into
account that the shell is well-defined in both X-ray and radio bands, we
considered a ring with inner and outer radii of 0.18 and 0.24 degrees, respectively,
rather than integrating the emission from the complete shell. The results are presented in
Fig.~\ref{fig:ext} and Fig.~\ref{fig:azprof} and indeed confirm the initial
observation of X-ray dimming towards the western part of the remnant.
Interestingly, the smoothed 325\,Mhz radio map seems to follow the same trend,
and in general exhibits a behaviour similar to that in X-rays. On the other hand,
the TeV emission shows no significant azimuthal variations with the western part
being one of the brightest parts of the shell.

\begin{figure*}[t]
    \centering
        \includegraphics[width=0.33\textwidth]{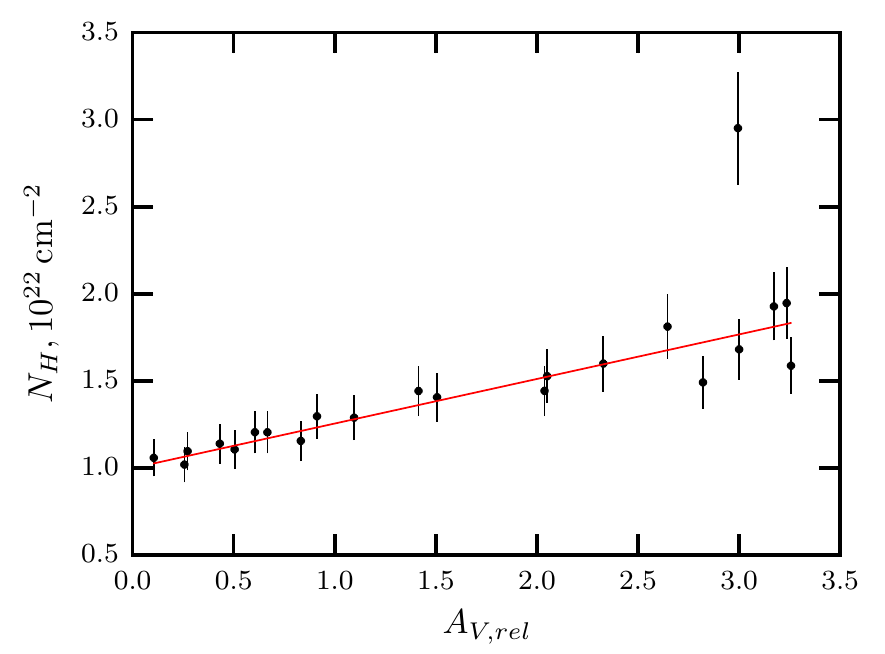}
        \includegraphics[width=0.33\textwidth]{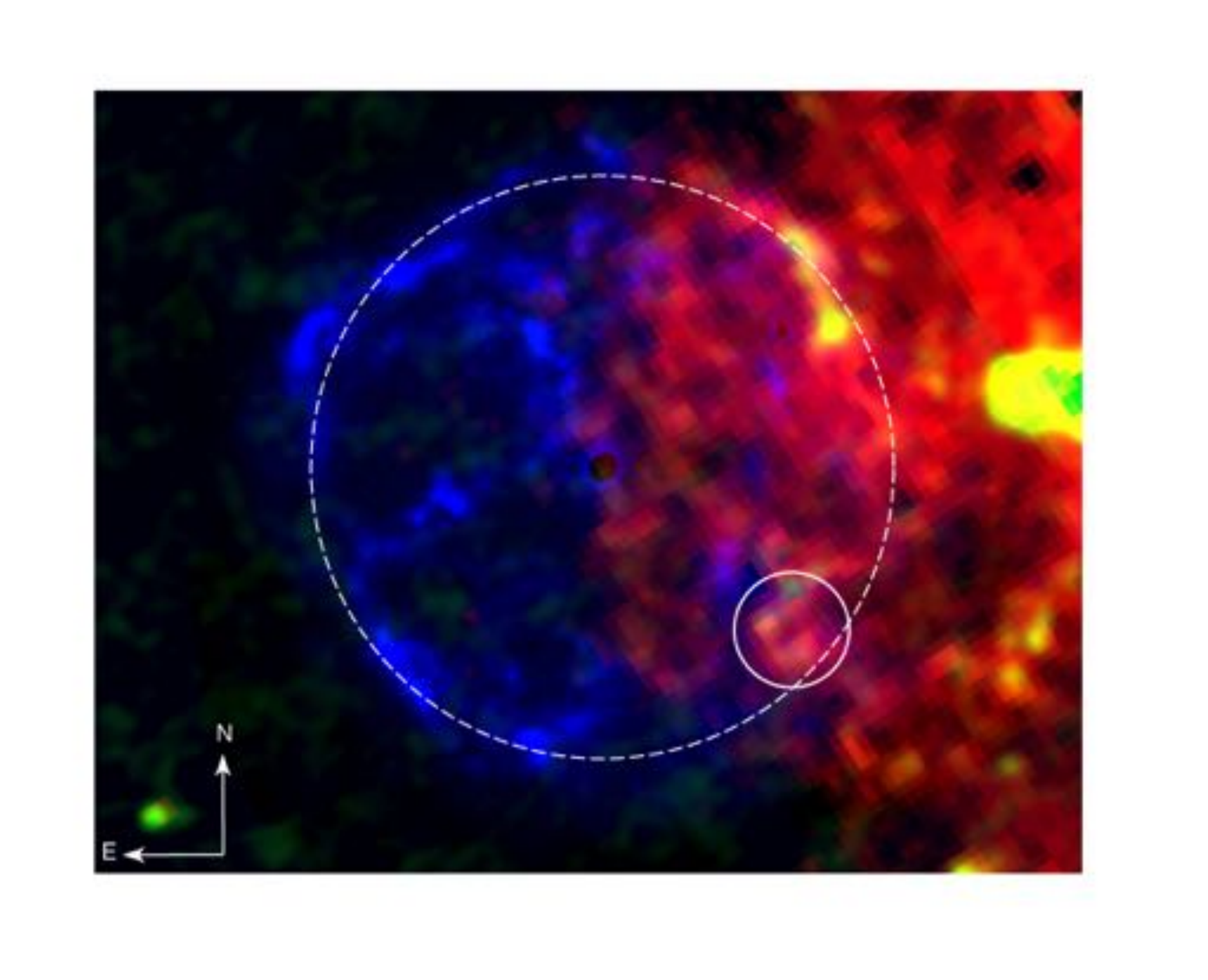}
        \includegraphics[width=0.33\textwidth]{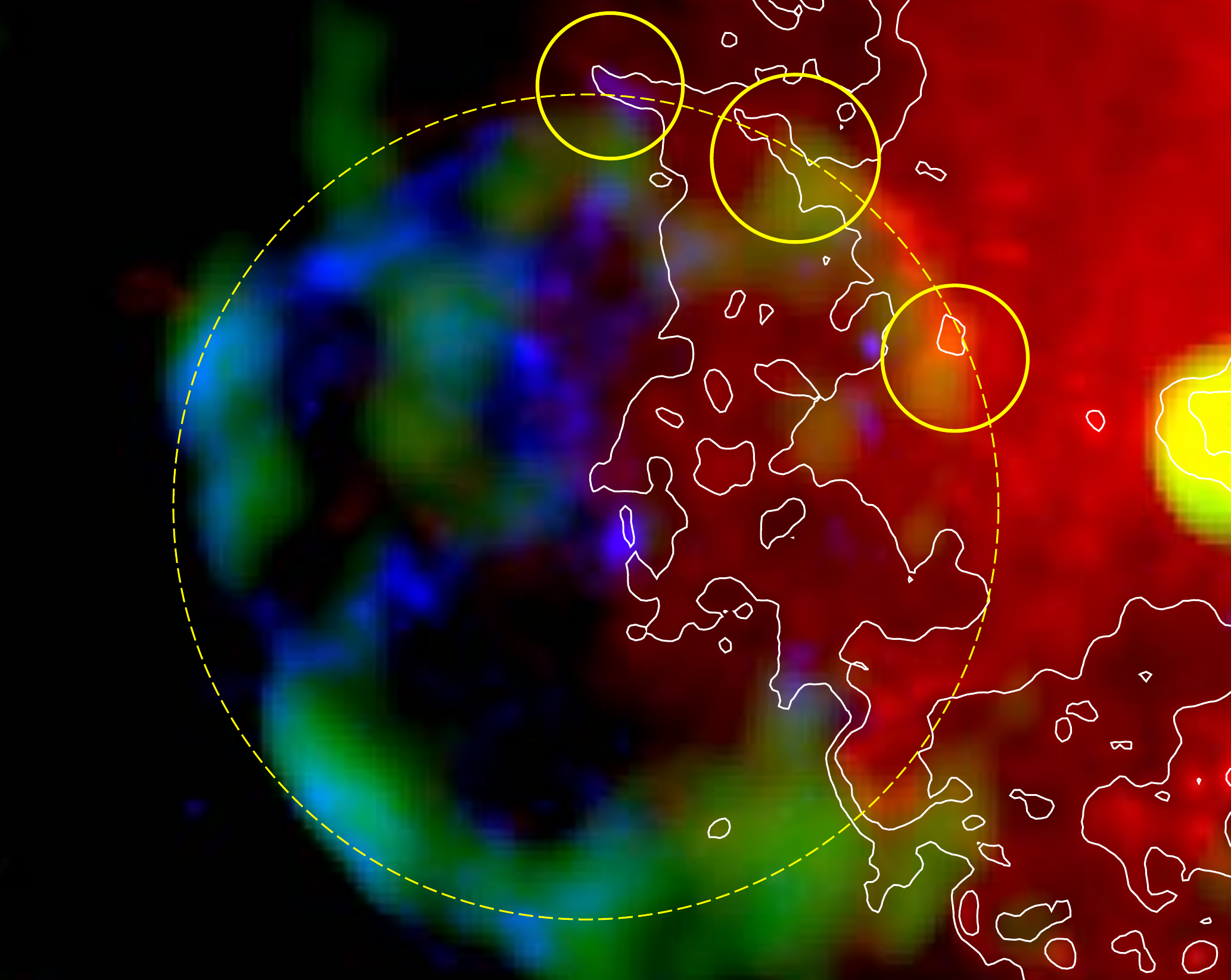}
    \caption{
    Correlation of the absorption column measured from the spatially resolved
    spectral analysis of \xmm data and the relative reddening derived from VVV
    photometry (left panel). The outlier point corresponds to region 2 where
    the optical extinction is underestimated due to the lack of stars in the optical
    data. We find $N_H\sim1+A_{V,rel}/4$ from a linear fit including a 10\%
    systematic uncertainty for the column measured in X-rays. A composite image
    of the calibrated absorption column density map derived from optical extinction
    (red) and ATLASGAL 0.87\,mm (green) maps is shown together with the X-ray 0.4-10\,keV
    image (blue) for reference in the middle panel. We note the enhanced absorption along
    the western rim of the remnant, particularly along its north-western part.
    The same features can be traced in the composite image showing X-ray (blue), radio (green), and 350\,$\mu$m \emph{Herschel}/SPIRE data (right panel).
    The white contours correspond to \emph{Herschel} data to highlight the extension of the filament in
    the north-west in northern direction where
    it bridges with the X-ray filament as shown with the northernmost yellow circle (see also the middle panel where
    the X-ray filament is more clearly seen). The other two circles emphasise
    the correspondence between the infrared and radio data along the filament.}
    \label{fig:extinction}
\end{figure*}
\subsection{Distance and association with the molecular cloud}
\label{sec:dist_mc}
As already mentioned, a rather robust lower limit on the distance can be obtained,
due to the spatial correlation of the X-ray absorption and the $^{12}CO$ emission from
material located in the Scutum-Crux arm at $\sim3$\,kpc. On the other hand,
the observed TeV flux implies that the distance to the source cannot be much
larger than that (i.e. not behind the Galactic centre). Most star forming
regions trace the Galactic spiral arms, so this immediately triggered the idea
that the remnant might be in the same arm and interact with the foreground
molecular cloud. To confirm this hypothesis, however, an
independent distance estimate to the source was needed, which proved to be more difficult.

The analysis of the X-ray spectrum of the central compact object suggests that
distances larger than $\sim5$\,kpc imply uncomfortably large NS radii for the
CCO in the centre of the remnant \citep{Klochkov13,Klochkov15}, so the remnant
most likely resides either in the Scutum-Crux ($\sim3$\,kpc) or in the Norma-Cygnus
($\sim4.5$\,kpc) Galactic arms.

The distance to the source can also be estimated by modelling the SED of a
bright central star illuminating the recently discovered dust shell within the
remnant that is also associated with the CCO \citep{Doroshenko16}. In
particular, for standard dust composition \cite{Vickers15} estimate a distance
of 3.8(7)\,kpc, i.e.  close to the lower limit obtained above.

This estimate is barely consistent with the distance estimate of $\sim5.2$\,kpc
obtained by \cite{Fukuda14} based primarily on a qualitative comparison of the
morphology of the TeV and HI radio emission. We note, however, that
\cite{Maxted15}, using the higher resolution MOPRA data, detected no
significant CS emission around $\upsilon_{\rm LSR}=-85$\,km\,s$^{-1}$ suggested
by \cite{Fukuda14}. On the other hand, \cite{Maxted15} were able to detect
several dense clumps of matter at $\upsilon_{\rm LSR}\sim-15$\,km\,s$^{-1}$
(corresponding to a distance of $\sim3$\,kpc) seemingly aligned with the SNR
shell as observed in radio continuum. It is interesting to note that this
velocity is close to the peak in $^{12}CO$ emission at $\sim-18$\,km\,s$^{-1}$,
which, as discussed above, is associated with the bulk of matter responsible
for the X-ray absorption. 

Higher resolution MOPRA maps presented by \cite{Maxted15} show that the
velocity of the denser clumps tracing the western rim of the SNR deviates
slightly from that for the bulk of the surrounding material. A similar behaviour
is, in fact, also observed  in other interacting SNRs
\citep{Moriguchi05,Castelletti13}, and thus can be viewed as an argument for
the physical association of \src with the molecular cloud at
$\sim-18$\,km\,s$^{-1}$. The absence of such peculiarities in the velocity
of the nearby material was quoted by \cite{Nayana17} as an argument against the interaction
of the SNR with nearby molecular clouds, which thus does not hold.
It should be noted, however, that this argument was motivated by  \cite{Dame01} data
which has limited spatial resolution, so it would be hard to notice any peculiarities in the first place.
The main issue with this argument is that the mechanisms responsible for acceleration of clumps are
unclear, and that the clumps themselves are rather compact, so there is a possibility
of a random spatial coincidence with the shell. This possibility could be ruled
out with more sensitive and higher resolution radio observations which would  more accurately trace the western rim of the SNR both in lines and continuum.

Here we add that to some extent the alignment of the dense clumps with the SNR
shell can be confirmed by an analysis of data from other wavelengths. In
particular, the dense clumps traced by CS emission can also be observed as a
thin dark filament in Spitzer MIPS/IRAC data \citep[][see
Fig.~1]{Doroshenko16}, and in submillimetre emission in the ATLASGAL survey
\citep{atlasgal} as illustrated in Fig.~\ref{fig:ext}. Moreover, the same
structure can also be detected in optical extinction maps where it can be
traced along the entire western part of the remnant, and seems to be perfectly
aligned with the detected X-ray shell in the north-west. While this does not
unambiguously prove their association, such a coincidence seems unlikely to be
just by chance.

To estimate the column density from optical reddening we used the recently
released \emph{PNICER} package \citep{pnicer}, and photometry from the VISTA
VVV survey \citep{vvv}. The extinction map with an angular resolution of
$1^\prime$ was produced using the mid-infrared colours and the spectral extraction
region 1 as a reference field. The result was then calibrated to the neutral
column density using the absorption column densities measured across the
remnant using \xmm data as illustrated in Fig.~\ref{fig:extinction}. It is
important to emphasise that the X-ray absorption and optical extinction are
closely correlated, with the sole exception of region 2 where the extinction is
underestimated due to the strong absorption. The resulting absorption map is
closely correlated with the sub-mm emission coming from the most absorbed areas
(which also coincides with the CS emission in radio), and arguably better
illustrates the alignment of the absorbing material with the northern edge of the
remnant. The individual clumps detected in the CS and sub-mm bands also correspond 
to the strongest extinction; however, a less intense rim connecting them is also
visible in the extinction map. We note that the filament detected in CS/sub-mm in
regions 15-19 extends further to the north in the extinction map, and is
aligned with the X-ray shell.

It is interesting to note that in the infrared band, these filaments are
clearly observed to be in the foreground with respect to the bulk of the emitting material
\citep[see Fig.~1 in][]{Doroshenko16}, which is likely associated with the CO
emitting material and is thus  responsible for the X-ray absorption. The 3-D mapping of
the mid-infrared extinction by \cite{Schultheis14} seems to support this
observation, although the resolution of the published maps is probably insufficient
to draw any firm conclusions. Nevertheless, the larger filament in the south-west
marked with a circle in Fig.~\ref{fig:extinction} has a comparable size to and is
aligned with the filiment shown in the 3-D extinction map sky pixels presented by
\cite{Schultheis14}. As shown in Fig.~\ref{fig:3dext}, the distribution of
matter in this direction is similar to that within the entire remnant, which suggests
that the bulk of the absorbing material in the filament must be relatively close
to the larger cloud. However, the distribution seems to be slightly skewed to the front,
which could explain why the filaments appear dark in the infrared.

On the other hand, as concluded above, the filament seems to be aligned with
the western rim of the SNR and thus is  at the same distance as the
remnant itself. However, the remnant must be  in the background of the material
absorbing X-ray and optical emission. There is, therefore, an apparent
contradiction since the remnant and the filament cannot be in front of and behind
 the molecular cloud at the same time. This is possible  if the apparent
alignment is still a chance coincidence or if the remnant and the molecular
cloud responsible for the absorption of the observed X-ray emission are
actually at the same distance and interact with each other. In this case the
fact that part of the shell appears in the foreground could be explained by
irregularities in the distribution of material within the absorbing cloud.

This conclusion is supported by the fact that no significant X-ray emission is
detected from the western part of the remnant. Indeed, there is an apparent
difference in the X-ray morphology in the eastern and western parts which needs to be
explained. 
The easiest explanation is then that the SNR shock encountered dense material
in the west
and slowed to the point when synchrotron falls out of the X-ray band
(keeping in mind that radio emission is still observed from the region). This scenario
has in fact been  discussed by \cite{Cui16}, who suggested that the observed
X-ray synchrotron emission from the shell requires high shock velocities, which
can only be retained if the remnant expands into an essentially empty stellar
bubble blown by the progenitor. The absence of X-ray emission from the western part
suggests that the shock velocity is much lower there, pointing to interaction
with dense molecular clouds. If this is indeed the case, the interaction
must have started quite recently since otherwise the circular shape of the
radio and TeV shells would also have  been distorted. Furthermore, once the
interaction starts the strong shock must start heating the swept-up material,
eventually producing thermal X-ray emission which has not yet been detected.

We note that in the case of the SNR RX\,J1713.7$-$3946, which displays a similar 
large-scale X-ray intensity trend, a different scenario has been invoked. There,
the X-ray brightening has been interpreted as coming from an interaction
with molecular clouds, while the dimmer part has been associated with lower density
surrounding material \citet{Cassam04}. In the case of HESS\,J1731$-$347, 
this  interpretation is not supported by the available gas density tracers. 
Also, the TeV emission in the west of the SNR which does not follow the dimming 
trend would need to be explained by processes not directly related to the SNR. 
A more straightforward explanation is therefore opposite to RX\,J1713.7$-$3946, 
namely that the SNR shock encountered dense material in the X-ray dim regions. 

Alternatively, the absence of bright filaments in the western part could be
explained as being projection effects. However, this also requires interaction
with the inhomogeneous medium. We conclude, therefore, that the possibility that
the remnant interacts with the cloud is consistent with observations, and this region is thus a plausible
target to search for high energy hadronic emission. In particular, it would be important
to search for changes in the TeV spectrum in the western part of the remnant with additional VHE observations.
Indirect evidence can also be obtained based on deeper radio observations aimed to better constrain
suggested changes of the spectral index, and independently estimate the magnetic field as discussed by \cite{Nayana17}.

\section{Conclusions} \xmm observations of the SNR \src covered  the entire shell visible in TeV and radio bands for the first
time. In the X-ray band, the
morphology of the remnant is quite irregular:  the eastern part is
considerably brighter than the western part (which is closer to the Galactic
plane). The emission from the eastern part is dominated by narrow, well-defined
filaments correlated with radio emission, whereas the western part is dimmer and more diffuse.
A spatially resolved spectral analysis confirmed the previously reported
\citep{discovery_paper} non-thermal origin for the X-rays. The increase in
the absorption column towards the Galactic plane is correlated with the CO emission
observed in radio, which  puts a lower limit on the distance to the source
of $\sim3.2-4$\,kpc. This estimate is consistent with the observed velocities
of dense CS clumps reported by \cite{Maxted15}, which indeed seem to be tracing
the western rim of the shell.

Based on the updated X-ray and radio fluxes, we find that the broadband spectral
energy distribution is consistent with synchrotron in the radio to X-ray and
inverse Compton in the TeV bands, although the morphology of the shell in
the respective bands seems to be slightly different. In particular, the X-ray and
radio continuum emission is suppressed towards the Galactic plane, which is
not the case for the TeV emission where the shell shows a flat azimuthal profile.
Taking into account the CS emission and enhanced optical extinction along the
western part of the remnant, we conclude that the remnant is most likely
interacting with the nearby dense molecular cloud. This conclusion supports the
SNR evolution scenario outlined by \cite{Cui16}, and might signify that part of
the observed TeV emission actually has a hadronic origin.

Based on the spatially resolved spectral analysis of the extended emission we do
not find, however, any significant variations in the intrinsic X-ray spectrum of the
shell, which could be expected for a remnant interacting with the
non-uniform ISM. On the other hand, minor variations in the X-ray spectrum
still cannot be completely excluded due to the strong stray-light
contamination affecting the \xmm data. Furthermore, the brightness of the 
shell in the western part is not sufficient for a detailed spectral analysis.
The fact that it is so low can be interpreted as an argument
supporting the interaction of the shell with the cloud.
Finally, we also find no evidence for thermal emission from the ejecta or
reverse shock. The latter is not surprising given the amount of cold dust
observed within the shell.

\begin{acknowledgements}
The authors would like
to thank the Deutsches Zentrums für Luft- und Raumfahrt (DLR) and Deutsche
Forschungsgemeinschaft (DFG) for financial support (grants DLR~50~OR~0702, FKZ
50 OG 1301, SA2131/1-1, FKZ OR 1310). W.W. Tian acknowledges support from the NSFC
(11473038). A.B. acknowledges support from the   Grant-in-Aid for
Scientific Research of the Japanese Ministry of Education, Culture, Sports,
Science and Technology (MEXT) of Japan, No.15K05107. The authors  thank the anonymous referee
for the timely and useful comments which helped to improve the manuscript.
\end{acknowledgements}

\bibliography{biblio}
\end{document}